# Negative refraction in hyperbolic hetero-bicrystals

A. J. Sternbach[1], S. L. Moore[1], A. Rikhter[2], S. Zhang[1], R. Jing[1], Y. Shao[1], B. S. Y. Kim[3], S. Xu[1], S. Liu[3,4], J. H. Edgar[4], A. Rubio[5,6], C. Dean[1], J. Hone[3], M. M. Fogler[2] and D. N. Basov[1]

1. Department of Physics, Columbia University, New York, NY, USA

2. Department of Physics, University of California San Diego, San Diego, CA, USA

3. Department of Mechanical Engineering, Columbia University, New York, NY, USA

4. Tim Taylor Department of Chemical Engineering, Kansas State University, Manhattan, KS, USA

5. Center for Computational Quantum Physics (CCQ), Flatiron Institute, 162 Fifth Avenue, New York, New York 10010, USA

6. Max Planck Institute for the Structure and Dynamics of Matter, Luruper Chaussee 149, 22761 Hamburg, Germany

**We visualized negative refraction of phonon polaritons, which occurs at the interface between two natural crystals. The polaritons - hybrids of infrared photons and lattice vibrations - form collimated rays that display negative refraction when passing through a planar interface between the two hyperbolic van der Waals materials: molybdenum oxide (MoO$_3$) and isotopically pure hexagonal boron nitride (h$^{11}$BN). At a special frequency $\omega_0$, these rays can circulate along closed diamond-shaped trajectories. We have shown that polariton eigenmodes display regions of both positive and negative dispersion interrupted by multiple gaps that result from polaritonic level repulsion and strong coupling.**

Refraction is an elemental phenomenon in optics, in which a ray of light changes direction after traveling across an interface between two media (*1*). Refraction is considered "negative" if the refracted beam emerges on the same side of the interface normal as the incident one. This uncommon occurrence was demonstrated in artificial metamaterials (*2*) and superlattices (*3*) whose permittivity $\varepsilon$ and permeability $\mu$ are simultaneously negative. Negative refraction alters light amplification and emission (*4, 5*) as well as non-linear optics (*6*) and may also cause trapped light (*7, 8*) as well as "perfect" lensing (*9*). Interfaces between anisotropic meta-structures with rotationally misaligned principal axes can also enable negative refraction (*10-12*). Extreme anisotropy is offered by hyperbolic materials (HMs), whose hybrid light-matter modes – polaritons – are predicted to exhibit all-angle negative refraction at carefully crafted interfaces (*11, 13*). In this work, we studied polaritons in a previously unexplored class of hyperbolic hetero-bicrystals made of two thin crystals, molybdenum oxide (MoO$_3$) (*14-18*) and isotopically pure hexagonal boron nitride (h$^{11}$BN) (*19-21*). Our hyperspectral nano-imaging data reveal localization, negative refraction, and closed-loop circulation of polaritonic rays inside h$^{11}$BN-MoO$_3$ hetero-bicrystals. Central to the observed effects is the gap in the polaritonic dispersion, which we extracted from hyperspectral images of polaritonic waves.

The hyperbolic electrodynamics of both h$^{11}$BN (crystal A) and MoO$_3$ (crystal B) is born out of strong dipole active phonons (*22*). These resonances drive the permittivity negative along at least one principal axis, whereas positive "dielectric-like" positive permittivity is preserved along the remaining principal direction(s). Our results can be understood by focusing on the $x - z$ plane (Fig. 1) for frequencies at which the phonon (Reststrahlen) bands of the constituent crystals overlap, $740 \text{ cm}^{-1} < \omega < 822 \text{ cm}^{-1}$. At these frequencies, the permittivity of h$^{11}$BN is positive along $\hat{x}$ and negative along $\hat{z}$, $\varepsilon_A^x(\omega) > 0$, and $\varepsilon_A^z(\omega) < 0$ (type-I hyperbolicity). In MoO$_3$, the signs are reversed, $\varepsilon_B^x(\omega) < 0$ and $\varepsilon_B^z(\omega) > 0$ (type-II hyperbolicity; Fig. 1A) in the same frequency range.

It is customary to refer to electromagnetic modes of polar materials as polaritons. The polariton dispersion assumes a simple form $(q_x^2/\varepsilon^z) + (q_z^2/\varepsilon^x) = \omega^2/c^2$ when the polariton momentum $\vec{q} = (q_x, q_y, q_z)$ is in the $x - z$ plane, $q_y = 0$. In HMs, the polariton isofrequency lines are hyperbolas (Fig. 1B) (*14, 19, 20, 23*). The asymptotes of these hyperbolas are inclined by the angle $\pm\theta$ with respect to the $x$-axis, where $\theta = \theta(\omega)$, defined by $\tan\theta = i \sqrt{\varepsilon^x}/\sqrt{\varepsilon^z}$, is positive for type-I and negative for type-II HMs. In the high-$q$ limit, probed in our near-field experiments, the polariton group velocity $\vec{v} = \nabla_{\vec{q}} \omega$ becomes orthogonal to $\vec{q}$ (*24*). Because the angles $\theta_A > 0$ and $\theta_B < 0$ have opposite signs while momentum $q_x$ is conserved, the tangential velocity $v_x = -|v| \text{ sgn } q_x \sin\theta$ changes sign in refraction at the A-B interface. The net effect is that polaritons exhibit negative refraction (supplementary text, section S1).

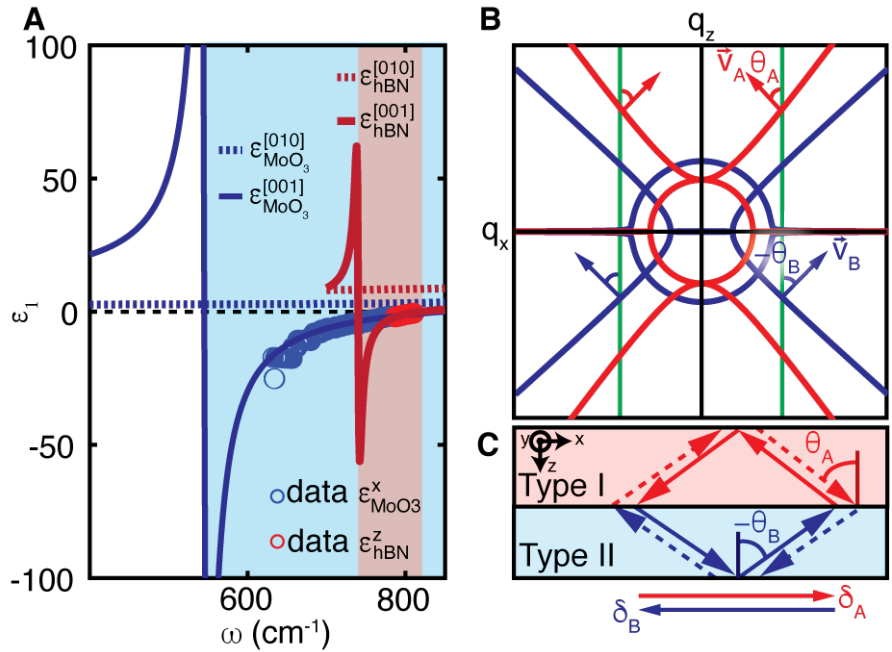

Figure 1| **Polaritons in hyperbolic hetero-bicrystals. A**, Real components of the permittivity, $\varepsilon_1$, of h$^{11}$BN and MoO$_3$. The dots are experimental data. The parameters for the calculations, indicated with solid lines, are extracted from our data (Table S1). **B**, Schematic showing q$_y$=0 cuts of the polariton isofrequency surfaces of type-I (crystal A, red) and

type-II (crystal B, blue) HMs (supplementary text, section S1). The group velocities $\vec{v}_{A,B}$ and their tilt angles $\theta_A > 0$, $\theta_B < 0$ are indicated. **C**, Schematic of the polariton rays in a bicrystal assembled from a type-I HM (crystal A, h$^{11}$BN) and a type-II HM (crystal B, MoO$_3$). The lateral shifts inside the crystals $\delta_A > 0$, $\delta_B < 0$ are indicated with arrows. The ray paths are closed if $\delta_A + \delta_B = 0$.

We report on a new class of hyperbolic hetero-bicrystal structures that reveal negative refraction of polaritons. If a hyperbolic ray emerges on the B-side of the A-B interface the ray will be laterally displaced by a distance $\delta_B/2 < 0$ after propagating through Crystal B. Negative refraction occurs at the interface with crystal A prompting an additional displacement $\delta_A/2 > 0$. At a frequency $\omega_0$ where the condition $\delta_A(\omega_0) + \delta_B(\omega_0) = 0$ is satisfied, polaritons travel in closed trajectories. Experimental signatures of the closed-cycle electrodynamics near $\omega_0$ are evident in our data (Figs. 2 and 3). However, these observations cannot be explained by polaritonic ray optics alone. We have shown that the principal modes of crystals A and B hybridize into a single strongly coupled eigenmode at $\omega_0$, leading to prominent gaps in frequency-momentum dispersion.

To visualize polaritons we used scanning near-field optical microscopy (SNOM). In SNOM measurements the metalized tip of an atomic force microscope probes optical effects with sub-diffractional spatial resolution, roughly given by the tip's radius, which is about 20 nm (*44*). To meet the demand for quasi-monochromatic excitation at frequencies within the overlapping Reststrahlen bands of h$^{11}$BN and MoO$_3$ (Fig. 1A) (*25*) we generated ultra-narrowband mid-infrared pulses with the spectral bandwidth <4 cm$^{-1}$ (supplementary text, section S2.5).

Nano-imaging data unequivocally demonstrated negative refraction in h$^{11}$BN/MoO$_3$ hetero-bicrystals (Fig. 2). We patterned a gold strip with a width of $2w \approx 750$ nm on the surface of silicon dioxide (SiO$_2$). The sharp edges of the strip along the y-axis enhance the infrared field and excite polaritons in the bicrystal with $q_y \approx 0$ (*26*). A MoO$_3$ crystal was placed on top of the launcher with its *c*-axis perpendicular to the strip (supplementary Fig. S7). We obtained images of the scattering amplitude, $|s|$, at the temperature T=99 K to minimize losses. Images of $|s|$, collected at the surface of MoO$_3$ (Fig. 2B), reveal a pair of characteristic twin-peak profiles near the edges of the launching strip (Fig. 2C, inset: marked 1, 2, 3, and 4). The separation, $\delta_B$, between peaks 1&2, or equivalently 3&4, is consistent with the directional propagation of hyperbolic rays introduced in Fig. 1 (*14*). Further, the magnitude of $\delta_B$ increases as the infrared frequency decreases (Supplementary Fig. S8), which also supports the notion of conical ray propagation in MoO$_3$ that is characteristic for a hyperbolic medium.

Next, we placed a crystal of h$^{11}$BN on top of the MoO$_3$/Au (gold) assembly and visualized the nano-optical intensity at the top of the hetero-bicrystal. We observed a single peak of $|s|$ in relation to each edge of the Au strip at $\omega_0 = 787 \text{ cm}^{-1}$ (Fig. 2, A and C). We also detected a considerable intensity between the two peaks (supplementary text, sections S1, S2.6). Our observations, augmented with modeling, are consistent with negative refraction guiding the hyperbolic rays to the same lateral positions at the top and bottom surfaces of the bicrystal (Fig. 2C, top inset). Effectively, negative refraction delivers a projection of the Au strip to the top surface of the bicrystal through diverging and converging trajectories of the hyperbolic rays inside the bicrystal. Numerical simulations capture gross features of the data in Fig. 2, A and B (analysis of subtle differences between the model and experiments is provided in the supplementary text, section S1.3). The totality of data in Fig. 2 and Fig. S8 establish negative refraction at the h[11]BN-MoO$_3$ interface.

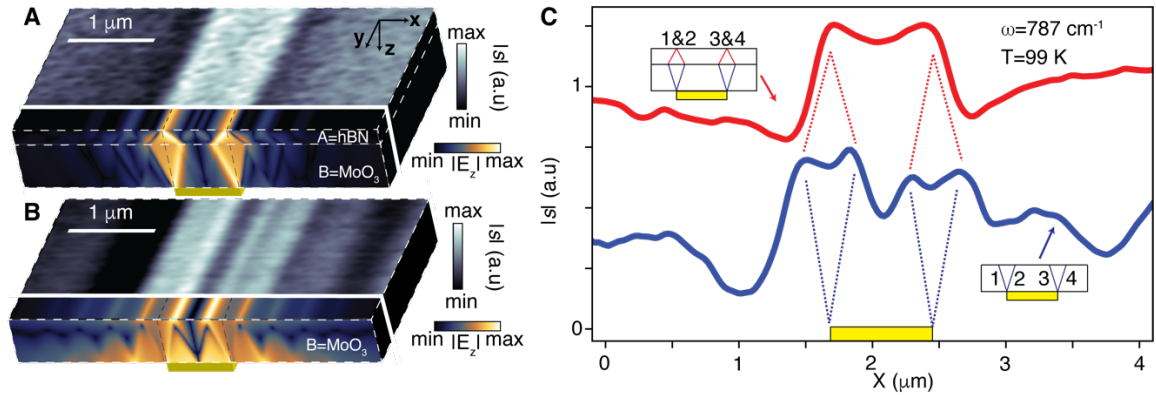

Figure 2| **Negative refraction of polaritons. A-C**, Near-field amplitude data, $|s|$, obtained at various surfaces in the x – y plane of an h[11]BN/MoO$_3$/Au stack. All data were obtained with at ω=787 cm$^{-1}$ at temperature T=99K, with thicknesses d$_{hBN}$=98 nm on d$_{MoO3}$=290 nm A=h[11]BN and B=MoO$_3$ crystals, respectively. **A**, Imaging data of $|s|$ in perspective at the top surface of h[11]BN-MoO$_3$-Au. **B**, Data obtained at the surface of MoO$_3$-Au, displayed in an identical manner to that of panel (**B**). Calculations of $|E_z|$ in the xz-plane, and a strip in the xy-plane, are also shown in false color (supplementary text, section S1) in (**A**) and (**B**). Yellow rectangles indicate gold bars, beneath the HMs, and black dashed lines indicate the strip's edges. **C**, Line profiles of $|s|$ as a function of the real-space coordinate, X. (Insets) The geometry in the x – z plane. Two pairs of hyperbolic rays, 1&2 and 3&4, launched by the two edges of the Au strip are labeled.

We then inquired into the frequency-momentum ($\omega, q_x$) dispersion of the hetero-bicrystal polaritons and its implications for the observed negative refraction. We collected hyperspectral data of the frequency dependent near-field amplitude $|\tilde{S}(X,\omega)|$ as a function of the distance $X$ from the bicrystal edge, following established procedures (*20, 27, 44*). Except for a narrow window of frequencies around $\omega_0 = 787 \text{ cm}^{-1}$, we witnessed oscillations (or fringes) of $|\tilde{S}(X,\omega)|$ in our hyperspectral data (Fig. 3B). The period of the oscillations identified in Fig. 3B systematically varies with $\omega$. Thus, our observations revealed how the wavelength of polaritonic waves, $\lambda_p(\omega)$, evolves with the frequency of incident infrared light. The data in Fig. 3B, provide access to the polaritonic ($\omega, |q_x|$) dispersion, because $\lambda_p(\omega) = 2\pi/|q_x(\omega)|$ (Fig. 3C). We stress a non-monotonic trend of $\lambda_p(\omega)$. Indeed, $\lambda_p(\omega)$ decreased when the frequency was near the lower bound of the overlapping Reststrahlen bands, but then reversed the trend near the upper bound of this frequency range. Near the frequency $\omega_- = 773 \text{ cm}^{-1}$, we detected two different fringe periods; hence, there are two sets of $q_x$ points in the vicinity of $\omega_-$ in Fig. 3C (Fig. S6). These features, at $\omega_0$ and $\omega_-$, are not present in the dispersions of constituent crystals (Fig. S11). Thus, the hyperspectral data in Fig.3 indicate that polaritons in the bicrystal are coupled modes.

A standard method for calculating the polariton dispersion involves finding the maxima of the reflection coefficient $r_p = r_p(\omega, |q_x|)$ of a $p$-polarized plane wave (*20, 27-29*). The results for the imaginary part of the p-polarized reflection coefficient (Im $r_p$) (Fig. 3) reveal the existence of multiple dispersion branches. The data points match the calculated branches with the smallest $|q_x|$, the so-called principal modes. The full dispersion of the bicrystal displays a nonmonotonic $|q_x(\omega)|$ punctuated by spectral gaps (Fig. 3C). This dispersion can be understood as the family of avoided crossings exhibited by the modes of the constituent crystals. The polariton branches have a negative dispersion in crystal A (Fig. 3D, red curves) and positive dispersion in Crystal B (Fig. 3D, blue curves; see also Fig. S11). Accordingly, the dispersion of the coupled modes of the bicrystal alternates in sign each time $|q_x|$ passes through an avoided crossing. The locations of the crossings are determined by a Bohr-Sommerfeld-like quantization condition:

$$(\delta_A + \delta_B)q_x = \pi n + \text{const,} \tag{1}$$

where $n$ is an integer. Equation (1) implies that the frequency $\omega_0$, at which $\delta_A + \delta_B$ vanishes, is typically gapped at all $q_x$, which agrees with Fig. 3C. Our modeling predicts that the magnitude of these gaps scale with the polariton's velocity. Therefore, the gap decreases as $\sim 1/|q_x|$, at large $|q_x|$ (supplementary text, section S1). Within the gaps, the pole of $r_p(\omega, q_x)$ occurs at a complex $q_x$ with a nonzero imaginary part even in the absence of dissipation. Thus, exactly at $\omega_0$ the polaritonic modes are evanescent, i.e., exponentially localized near a launcher because of the combined effects of negative refraction and wave interference. We observe a gap near $\omega_0$ (Figs. 3C) situated at $|q_x| = 24\ \mu m^{-1}$ with the size $\Delta\omega = 13 \pm 3\ cm^{-1}$, which is in good agreement with the calculated value of $\Delta\omega(|q_x| = 26\ \mu m^{-1}) = 16\ cm^{-1}$ (Figs. 3, C and D). The hetero-bicrystal polaritons visualized here comply with the definition of the strong mode coupling: The magnitude of the gap exceeds the linewidth of the mode (supplementary text, section S1).

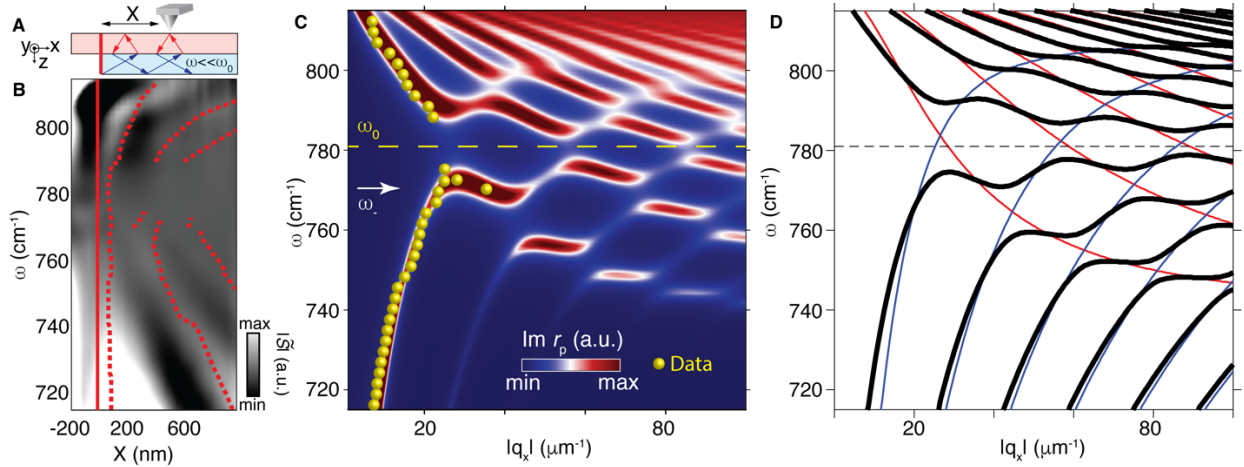

Figure 3| **Spectral gaps in the hetero-bicrystal dispersion.** All data were obtained on a h[11]BN/MoO$_3$ bicrystal with thicknesses of d$_{hBN}$=58 nm and d$_{MoO3}$=150 nm at ambient temperature. **A**, Schematic illustrating ray trajectories in h[11]BN (red) and MoO$_3$ (blue) for $\omega \ll \omega_0$. **B**, Amplitude data, $|\tilde{S}(X,\omega)|$ as a function of the distance, X, between the tip and bicrystal edge (solid red line). The edge of h[11]BN is located at X = -700 nm (supplementary text, section S1). The red dashed lines indicate locations where maxima are observed in our calculations (Fig. S6D). **C**, The imaginary part

of the p-polarized reflection coefficient (Im $r_p$) is shown as a function of $\omega$ and the absolute value of the momentum component, $|q_x|$. The calculation uses realistic room-temperature losses of h$^{11}$BN and MoO$_3$ (table S1). Data points are indicated with yellow dots (Fig. S6). **D**, The bicrystal dispersion is indicated with black lines for the idealized case with vanishing losses. Thin color traces indicate the dispersions the parent crystals, MoO$_3$ (blue), and h$^{11}$BN (red) calculated by using parameters in table S1.

In this work, we introduced hyperbolic hetero-bicrystal polaritons. We showed that the interface polaritons in h$^{11}$BN-MoO$_3$ can display negative refraction, spectral gaps, strong coupling, and localization. These attributes of hetero-bicrystals are broadly relevant to photonic applications (*30, 31*) by using HMs. Moreover, polaritons in hetero-bicrystals can be focused to sub-diffraction-limited spot-sizes (*18, 32, 33*), which can enable perfect lensing by means of negative refraction (*9*). The attainable focal spots can, however, be limited by extrinsic factors, including crystal losses and imperfect polaritonic launchers (Fig. S15). Further, similar to Fabry-Pérot cavities, negative refraction can cause radiation to propagate in closed cycles in our hetero-bicrystal nano-cavities. Dielectric losses remain a challenge but could possibly be mitigated with active loss compensation (*4, 5, 34*).

# REFERENCES

1. N. Yu *et al.*, *Science* **334**, 333-337 (2011).
2. R. A. Shelby, D. R. Smith, S. Schultz, *Science* **292**, 77-79 (2001).
3. A. Pimenov, A. Loidl, P. Przyslupski, B. Dabrowski, *Physical Review Letters* **95**, 247009 (2005).
4. S. Xiao *et al.*, *Nature* **466**, 735-738 (2010).
5. T. Pickering, J. M. Hamm, A. F. Page, S. Wuestner, O. Hess, *Nature Communications* **5**, 4972 (2014).
6. A. K. Popov, V. M. Shalaev, *Opt. Lett.* **31**, 2169-2171 (2006).
7. K. L. Tsakmakidis, A. D. Boardman, O. Hess, *Nature* **450**, 397-401 (2007).
8. K. L. Tsakmakidis, O. Hess, R. W. Boyd, X. Zhang, *Science* **358**, eaan5196 (2017).
9. J. B. Pendry, *Physical Review Letters* **85**, 3966-3969 (2000).
10. Y. Zhang, B. Fluegel, A. *Physical Review Letters* **91**, 157404 (2003).
11. X. Lin *et al.*, *Proceedings of the National Academy of Sciences* **114**, 6717 (2017).
12. A. J. Hoffman *et al.*, *Nature Materials* **6**, 946-950 (2007).
13. J. Jiang, X. Lin, B. Zhang, *Research* **2018**, 2532819 (2018).
14. W. Ma *et al.*, *Nature* **562**, 557-562 (2018).
15. G. Álvarez-Pérez *et al.*, *Science Advances* **8**, eabp8486.
16. G. Hu *et al.*, *Nature* **582**, 209-213 (2020).
17. M. Chen *et al.*, *Nature Materials* **19**, 1307-1311 (2020).
18. J. Duan *et al.*, *Nature Communications* **12**, 4325 (2021).
19. A. J. Giles *et al.*, *Nature Materials* **17**, 134-139 (2018).
20. S. Dai *et al.*, *Science* **343**, 1125 (2014).
21. H. Herzig Sheinfux *et al.*, arXiv:2202.08611 (2022).
22. Q. Zhang *et al.*, *Nature* **597**, 187-195 (2021).
23. Z. Jacob *et al.*, *Applied Physics B* **100**, 215-218 (2010).
24. R. K. Fisher, R. W. Gould, *Physical Review Letters* **22**, 1093-1095 (1969).
25. A. J. Sternbach *et al.*, *Nature Communications* **11**, 3567 (2020).
26. P. Li *et al.*, *Nature Communications* **6**, 7507 (2015).
27. E. Yoxall *et al.*, *Nature Photonics* **9**, 674-678 (2015).
28. T. Low *et al.*, *Nature Materials* **16**, 182-194 (2017).
29. H. N. S. Krishnamoorthy, Z. Jacob, E. Narimanov, I. Kretzschmar, V. M. Menon, *Science* **336**, 205 (2012).
30. G. Lu *et al.*, *Nano Letters* **21**, 1831-1838 (2021).
31. Z. Zheng *et al.*, *Advanced Materials* **34**, 2104164 (2022).
32. Q. Chen *et al.*, *Photon. Res.* **9**, 1540-1549 (2021).
33. S. Wuestner, A. Pusch, K. L. Tsakmakidis, J. M. Hamm, O. Hess, *Physical Review Letters* **105**, 127401 (2010).
34. J. Nishida *et al.*, *Nature Communications* **13**, 1083 (2022).
35. J.-S. Wu, D. N. Basov, M. M. Fogler, *Physical Review B* **92**, 205430 (2015).

36. G. Álvarez-Pérez et al., *Advanced Materials* **32**, 1908176 (2020).
37. Z. Zheng et al., *Science Advances* **5**, eaav8690 (2019).
38. S. Dai et al., *Nature Nanotechnology* **10**, 682-686 (2015).
39. S. Dai et al., *Nature Communications* **6**, 6963 (2015).
40. M. A. Huber et al., *Nature Nanotechnology* **12**, 207-211 (2017).
41. A. J. Sternbach et al., *Science* **371**, 617 (2021).
42. D. Akinwande et al., *Nature* **573**, 507-518 (2019).
43. D. W. McCamant, P. Kukura, S. Yoon, R. A. Mathies, *Review of Scientific Instruments* **75**, 4971-4980 (2004).
44. Supplementary Materials

**Funding**: Research on polaritons in van der Waals materials is supported as part of Programmable Quantum Materials, an Energy Frontier Research Center funded by the U.S. Department of Energy (DOE), Office of Science, Basic Energy Sciences (BES), under award DE-SC0019443. Work on negative refraction is supported through the Vannevar Bush Faculty Fellow program ONR-VB: N00014-19-1-2630. DNB is Moore Investigator in Quantum Materials EPIQS GBMF9455. The development of novel nano-photonics methods is supported by DOE-BES grant DE-SC0018426. Support for the h$^{11}$BN crystal growth comes from the Office of Naval Research, award number N00014-20-1-2474.

**Author Contributions**: AJS conceived of the study. AJS recorded and analyzed the near-field data with assistance from SZ, SX, SM, and RJ. SM, BK, and SL prepared the samples with guidance from JE, CD, and JH. YS performed far-field measurements. AR, and MF performed theoretical calculations with assistance from AJS. AJS, DNB, and MF wrote the manuscript with input from all the coauthors.

**Competing interests**: The authors have no competing interests.

**Data and materials availability**: All the data are included in the manuscript or the Supplementary materials.

# Supplementary materials for "Negative refraction in hyperbolic hetero-bicrystals"

A. J. Sternbach[1], S. L. Moore[1], A. Rikhter[2], S. Zhang[1], R. Jing[1], Y. Shao[1], B. Kim[3], S. Xu[1], S. Liu[3,4], J. H. Edgar[4], A. Rubio[5,6], C. Dean[1], J. Hone[3], M. M. Fogler[2] and D. N. Basov[1]

1. Department of Physics, Columbia University, New York, NY, USA

2. Department of Physics, University of California San Diego, San Diego, CA, USA

3. Department of Mechanical Engineering, Columbia University, New York, NY, USA

4. Tim Taylor Department of Chemical Engineering, Kansas State University, Manhattan, KS, USA

5. Center for Computational Quantum Physics (CCQ), Flatiron Institute, 162 Fifth Avenue, New York, New York 10010, USA

6. Max Planck Institute for the Structure and Dynamics of Matter, Luruper Chaussee 149, 22761 Hamburg, Germany

## Materials and Methods

Experimental Setup

We used the pseudo heterodyne technique (*14, 19, 20*) to extract the amplitude ($S$) and phase ($\varphi$) of the near-field signal in imaging experiments (Fig. 2) and nano-FTIR (*20, 27, 35*) to extract spectra of $S(\omega)$ and $\varphi(\omega)$ (Fig. 3) with spatial resolution approximately given by tip's radius of curvature, which is around 20 nm. In this work, we demonstrate a homebuilt monochromator for quasi-monochromatic nano-imaging of polaritons with a pulsed light source (*25*). Details of our experimental apparatus are in Supplementary section S2.5.

## Supplementary Text

Table of Contents:

## S1. Hyperbolic modes and hyperbolic rays

*S1.1 Bulk modes and their reflection/refraction at interfaces*

The dispersion of electromagnetic modes in an anisotropic crystal is governed by Fresnel's equation

$$\begin{aligned}
&\left(\varepsilon^x q_x^2 + \varepsilon^y q_y^2 + \varepsilon^z q_z^2\right)\left(q_x^2 + q_y^2 + q_z^2\right) \\
&\quad - \left[(\varepsilon^y + \varepsilon^z)\varepsilon^x q_x^2 + (\varepsilon^x + \varepsilon^z)\varepsilon^y q_y^2 + (\varepsilon^x + \varepsilon^y)\varepsilon^z q_z^2\right] q_0^2 \\
&\quad + \varepsilon^x \varepsilon^y \varepsilon^z q_0^4 \\
&= 0\,,
\end{aligned} \quad \text{(S1)}$$

where $\varepsilon^i = \varepsilon^i(\omega)$ are the dielectric permittivities along the principal axes of the crystal $i = x, y, z$, vector $\vec{q} = (q_x, q_y, q_z)$ is the mode momentum, and $q_0 = \omega/c$ is the free-space photon momentum. For our near-field experiments, the quasi-static limit $q^2 \gg q_0^2$ is pertinent. Here $q = \sqrt{q_x^2 + q_y^2 + q_z^2}$ is the magnitude of vector $\vec{q}$.

For a given direction of the $\vec{q}$ vector, Eq. (S1) has two solutions for $q$, which are referred to as the ordinary and extraordinary. To understand their basic properties, we can consider the case where $\vec{q}$ is in the $x$-$z$ plane, so that $q_y = 0$, which is relevant for our experimental setup. It is easy to show that in this case the dispersion equation for the ordinary mode simplifies to $q^2 = q_x^2 + q_z^2 = \varepsilon^y q_0^2$. This equation however has no real solutions that belong to the quasi-static range $q \gg q_0$ (unless $\varepsilon^y$ is unusually large and positive). Therefore, the ordinary modes play little role in our experiments. On the other hand, the dispersion of the extraordinary modes is

$$\frac{q_x^2}{\varepsilon^z} + \frac{q_z^2}{\varepsilon^x} = \frac{\omega^2}{c^2} \qquad (S2)$$

for $q_y = 0$. This equation does have solutions with large real momenta $q \gg q_0$ if $\varepsilon^z$ and $\varepsilon^x$ are of opposite sign. In this case, for a given fixed $\omega$, Eq. (S2) describes a hyperbola in the $q_x$-$q_z$ plane, see Fig. 1B of the main text and Fig. S1. The crystal is characterized as a hyperbolic material (HM) of type I if $\varepsilon^z < 0 < \varepsilon^x$ and a HM of type II if $\varepsilon^x < 0 < \varepsilon^z$. We call the extraordinary modes of such HMs hyperbolic polaritons.

At $q \gg q_0$, the hyperbolic polariton dispersion curve becomes asymptotic to the pair of straight lines

$$q_z = \pm q_x \tan\theta, \qquad \tan\theta = i\frac{\sqrt{\varepsilon^x}}{\sqrt{\varepsilon^z}} \qquad (S3)$$

tilted by the angle $\pm\theta$ with respect to the $x$-$y$ plane and the complementary angle $\theta^* = (\pi/2) - |\theta|$ with respect to the $z$-axis (Fig. S1). The angle $\theta = \theta(\omega)$ is an important parameter in the problem. It was introduced in the main text.

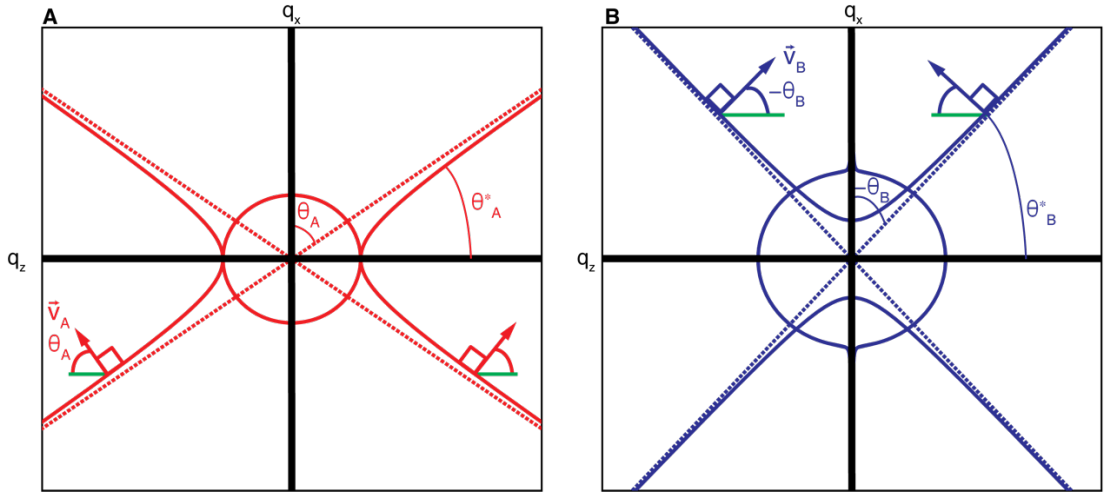

Figure S1| **Isofrequency curves of a hyperbolic crystal. A.** The $q_y = 0$ cross-section of the isofrequency surface in a type I HM with $\varepsilon^z < 0 < \varepsilon^x$ and $\varepsilon^y > 0$ consists of a hyperbola representing the extraordinary mode and an ellipse corresponding to the ordinary mode. The momenta of high-$q$ extraordinary modes are tilted by the angle $\theta^*$ with respect to the $z$-axis and the angle $\theta > 0$ with respect to the x-axis. The wavepackets of these modes form hyperbolic rays that propagate along the direction of the group velocity $\vec{v}_g$. The group velocity points away from the asymptotes (straight dashed lines). **B.** A similar diagram for a type II HM with $\varepsilon^x < 0 < \varepsilon^z$ and $\varepsilon^y > 0$ where the hyperbola appears on the other side of the asymptotes and $\theta < 0$, $\theta^*$ is also shown.

In a uniaxial HM (such as h$^{11}$BN) where $\varepsilon^y = \varepsilon^x$, Eqs. (S2) and (S3) are valid for an arbitrary $q_y$ if $q_x^2$ is replaced by $q_x^2 + q_y^2$. Hence, the isofrequency surface in a 3D $q$-space is a hyperboloid asymptotic to a circular cone with the opening angle $\theta$ in the momentum space. For MoO$_3$, which is biaxial, we have a distorted cone instead.

The polariton modes we discussed so far are plane waves of constant amplitude and fixed momentum $\vec{q}$. Such modes can be excited only by sources of infinite size. On the other hand, the field generated by a localized source of small size $a \ll q_0^{-1}$ has a strikingly different real-space structure in which the amplitude of the field is not constant but concentrated on a certain conical surface. It is useful to think that such a field configuration is composed of numerous hyperbolic rays. Each ray is a wavepacket of polariton modes centered around a given momentum $\vec{q}$. The rays propagate with the group velocity

$$\vec{v} = \nabla_{\vec{q}} \omega \,. \tag{S4}$$

From Eq. (S2), we find that for rays with momenta $q \sim a^{-1} \gg q_0$, this velocity is given by

$$v_x \simeq -\frac{\Omega \tan \theta}{q_x} \,, \qquad v_z \simeq \frac{\Omega \tan \theta}{q_z} \,, \qquad \Omega(\omega) \equiv \left(\frac{d}{d\omega} \tan \theta\right)^{-1} > 0 \,. \tag{S5}$$

(Although nonuniversal, the last inequality, i.e., $\Omega > 0$, is valid for both h$^{11}$BN and MoO$_3$ and should apply for most other HMs as well.) Equation (S5) indicates that the hyperbolic rays propagate at an angle $\theta$ with respect to the $z$-axis. This angle does not depend on the momentum: it is common for all the rays in the quasistatic limit. Therefore, the real-space field distribution created by the rays emitted by a localized source in an infinite HM is peaked at the surface of the so-called "resonant cone" inclined by the same angle $\theta$ from the $z$-axis. The $x$-$z$ cross-section of this resonant cone consists of two lines tilted by the angle $\pm\theta$ with respect to the $z$-axis. The relation between the momentum space and the real space distributions is reciprocal: a narrow isofrequency cone in $q$-space corresponds to a wide resonant cone in real space, and vice versa. The described properties can be also understood geometrically. Per Eq. (S4), $\vec{v}$ is normal to the polariton isofrequency surface, which is hyperboloidal or approximately conical. Hence, the momentum and the group velocity vectors of high-$q$ hyperbolic modes are nearly mutually orthogonal, $\vec{v} \cdot \vec{q} \approx 0$, see Fig. S1.

A key innovation of the heterostructures studied in our experiments is the interface of two different HMs: h$^{11}$BN (crystal A) and MoO$_3$ (crystal B). The hyperbolic modes experience reflection and refraction at this interface, which is oriented normal to the $z$-axis in our case. For either reflection or refraction, the sign and magnitude of the in-plane momentum component $q_x$ are conserved along the normal to the boundary. The angle between $q_x$ and the normal to the boundary is given by the complementary angle $\theta^*$ defined above for the geometry explored in this work. The direction of the reflected/refracted rays is determined by the group velocity. From Eq. (S5) we see that in a HM, the signs of $v_x$ and $q_x$ are not necessarily the same. This leads to the possibility of negative refraction (see Fig.1 of the main text for an illustration), where the refracted ray bends back rather than continues across the surface normal. Specifically, the group velocity component $v_x$ changes sign across the A-B interface if $\tan\theta_A$ and $\tan\theta_B$ have opposite signs. In other words, the negative refraction occurs if one of the crystals is of type I (so that $\tan\theta > 0$) and the other of type II ($\tan\theta < 0$). In contrast, the in-plane *phase velocity* component $(q_x/q)(\omega/q)$ has the same sign as $q_x$, and so it does not change sign across the interface.

Another process that occurs at the A-B interface is reflection. It is easy to show that under the assumptions we made previously ($q_x \gg q_0$ and $q_y = 0$), the reflection coefficient for the hyperbolic polariton incident on this interface from Crystal A has a momentum-independent value

$$r_{AB} = \frac{\bar{\varepsilon}_B - \bar{\varepsilon}_A}{\bar{\varepsilon}_B + \bar{\varepsilon}_A}, \qquad \bar{\varepsilon}_X \equiv \sqrt{\varepsilon_X^x}\sqrt{\varepsilon_X^z}. \tag{S6}$$

For the interface of two lossless HMs (with purely real $\varepsilon_j^z$ and $\varepsilon_j^x$), this quantity is real, and its magnitude $|r_{AB}|$ is between 0 and 1. This means that, in general, the polaritons are partially transmitted and partially reflected at the A-B interface. On the other hand, at the outer surface of an HM, i.e., at its interface with air, $\varepsilon^x = \varepsilon^z = \varepsilon_0$, the reflection coefficient, which we denote by $\beta$, is complex and its magnitude is equal to unity:

$$\beta = \frac{\bar{\varepsilon} - \varepsilon_0}{\bar{\varepsilon} + \varepsilon_0} = -e^{2i\pi\alpha(\varepsilon_0)}, \qquad \alpha(\epsilon) \equiv \frac{1}{\pi}\arctan\left(\frac{i\epsilon}{\bar{\varepsilon}}\right) - \frac{1}{2}. \tag{S7}$$

Therefore, at the outer surface, the polaritons experience a total internal reflection with the phase shift $\pi\alpha(\varepsilon_0)$. The reflection coefficient can be expressed in terms of these phase shifts as follows

$$r_{AB} = \frac{\beta_B - \beta_A}{1 - \beta_B\beta_A} = \frac{\sin[\pi\alpha_B(\varepsilon_0) - \pi\alpha_A(\varepsilon_0)]}{\sin[\pi\alpha_B(\varepsilon_0) + \pi\alpha_A(\varepsilon_0)]}, \tag{S8}$$

where the subscripts added to $\alpha$ and $\beta$ designate the material. In Fig. S2 we show the internal reflectivity $R_{AB}(\omega) = |r_{AB}|^2$ and the air reflectivity $R_{\text{air},X} \equiv |\beta_X|^2$ computed for the studied materials using optical constants from Table S1. Notice two items: i) the interval $740 \text{ cm}^{-1} < \omega < 822 \text{ cm}^{-1}$ of fractional reflectivity $0 \leq R_{AB} < 1$ where both crystals are hyperbolic and ii) a special frequency $\omega_{R0} = 814$ cm$^{-1}$ near the $z$-axis phonon frequency of h$^{11}$BN $\omega_{TO}^{[001]} = 822$ cm$^{-1}$ (see Table S1) where $\bar{\varepsilon}_B \approx \bar{\varepsilon}_A$, so that $R_{\text{air},A} \approx R_{\text{air},B}$ and, in accordance with Eq. (S8), the A-B interface is nearly reflectionless, $R_{AB} \approx 0$. As we will discuss in Sec. S1.2 below, smallness of $R_{AB}$ promotes coupling between the polariton eigenmodes of the two crystals.

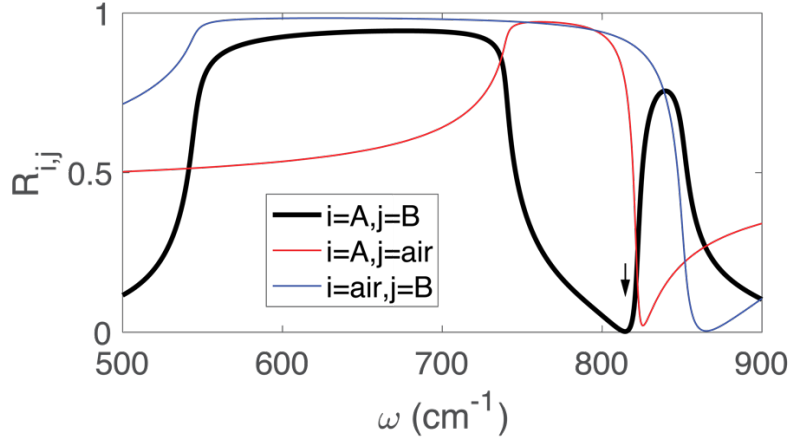

Figure S2 | **Large-momentum (near-field) reflectivity of material interfaces present in h$^{11}$BN/MoO$_3$ heterostructures.** The calculated reflectivity $R_{\text{air},A} = |\beta_A|^2$ of bulk crystal A (h$^{11}$BN) - air interface is shown by the red line, that of bulk crystal B (MoO$_3$) - air $R_{\text{air},B} = |\beta_B|^2$ is shown by the blue line, and that of the A-B interface $R_{AB}(\omega) = |r_{AB}|^2$ is represented by the black line. The frequency $\omega_{R0}$ where $R_{AB}$ vanishes is marked by the arrow. The tangential component of the polariton momentum is along the [001] principal axis of MoO$_3$.

## S1.2 Modes of layered heterostructures

A finite-thickness layered system can support "waveguide" polariton modes that are confined in the out-of-plane ($z$)-direction but have a definite momentum $(q_x, q_y)$ along the plane. Let us again assume that $q_y = 0$ so that $q = \sqrt{q_x^2 + q_y^2} = |q_x| \geq 0$. The momentum of the waveguide modes must be outside the light line, $q > q_0$. A standard method for finding the waveguide mode spectrum is to consider the reflection coefficient of a $p$-polarized plane wave incident on the surface of the sample from vacuum (or air). For a given frequency $\omega$, this coefficient $r_p = r_p(q, \omega)$ is a function of $q$. The waveguide modes show up as the poles of function $r_p(q, \omega)$ at complex $q = q' + iq''$ or equivalently, as sharp peaks of Im $r_p(q, \omega)$ at real and positive $q = q'$.

In turn, a standard approach for computing $r_p(q, \omega)$ is the transfer matrix method. For a heterostructure of $M$ layers indexed by $1 \leq j \leq M$ from top to bottom, this method leads to the following recursion relation (see, e.g., (36))

$$r_i = \frac{r_{i,i+1} + r_{i+1}\exp(2ik_{i+1}d_{i+1})}{1 + r_{i,i+1}r_{i+1}\exp(2ik_{i+1}d_{i+1})}, \qquad k_i = \sqrt{q_0^2 \varepsilon_i^x - \frac{\varepsilon_i^x}{\varepsilon_i^z} q^2} \qquad (S9)$$

for the reflection coefficient $r_i$ of a sub-stack of layers $i \leq j \leq M - 1$. By convention, layer index $j = 0$ refers to the air-filled half-space above the sample. The recursion is initialized at $i = M - 1$ with $r_{M-1} = r_{M-1,M}$ and is continued to successively smaller $i$. The desired reflection coefficient of the entire sample is $r_p = r_0$. Note that $k_i$ has the meaning of the $z$-axis momentum $q_z$ in the $i^{\text{th}}$ layer, cf. Eq. (S2), and that to compute the reflection coefficients $r_{i,i+1}$ of individual interfaces Eq. (S6) can be used.

Table S1 | **Optical constants of h¹¹BN and MoO₃.** Optical constants were obtained by fitting our experimental data in Fig. S11 and S13. These parameters are in a reasonable agreement with prior investigations of h¹¹BN (*19*) and MoO₃ (*37, 38*). For h¹¹BN, the [001] direction (c-axis) is along the $z$-axis, which is out of the plane. The [010] and [100] directions are along the $x$ and $y$-axis, respectively. For MoO₃, the [001] and [100] directions are in the $x$-$y$ plane, with the [001] axis making a $\varphi = 7.5^0$ angle with respect to the $x$-axis in the data of Figs. 3 and S11; the [010] direction is out of the plane, along the $z$-axis. Accordingly, the requisite components of the dielectric tensor of MoO₃ are given by $\varepsilon_B^x = \cos^2\varphi\, \varepsilon_B^{[001]} + \sin^2\varphi\, \varepsilon_B^{[100]}$, $\varepsilon_B^z = \varepsilon_B^{[010]}$. The '-' symbol indicates a quantity was not determined in this work.

| Parameter | Crystal A (h¹¹BN) | Crystal B (MoO₃) |
|---|---|---|
| $\varepsilon_\infty^{[100]}$ | 5.9 | 4.7 |
| $\omega_{LO}^{[100]}$ | 1608.7 [cm⁻¹] | 972 [cm⁻¹] |
| $\omega_{TO}^{[100]}$ | 1359.8 [cm⁻¹] | 820 [cm⁻¹] |
| $\Gamma^{[100]}$ | 2.5 [cm⁻¹] | 7 [cm⁻¹] |
| $\varepsilon_\infty^{[010]}$ | 5.9 | 2.4 |
| $\omega_{LO}^{[010]}$ | 1608.7 [cm⁻¹] | 1004 [cm⁻¹] |
| $\omega_{TO}^{[010]}$ | 1359.8 [cm⁻¹] | 958 [cm⁻¹] |
| $\Gamma^{[010]}$ | 2.5 [cm⁻¹] | – |
| $\varepsilon_\infty^{[001]}$ | 2.8 | 5.2 |
| $\omega_{LO}^{[001]}$ | 822 [cm⁻¹] | 851 [cm⁻¹] |
| $\omega_{TO}^{[001]}$ | 740 [cm⁻¹] | 545 [cm⁻¹] |
| $\Gamma^{[001]}$ | 4 [cm⁻¹] | 7 [cm⁻¹] |

For the electric field inside the layers, we derived the following recursion relations:

$$E_x(x,z) = -\frac{\partial}{\partial x}\Phi(x,z), \quad E_z(x,z) = -\frac{\partial}{\partial z}\Phi(x,z),$$

$$\Phi(x,z) = \int \frac{dq}{2\pi} e^{iqx} \widetilde{\Phi}(q,z),$$

$$\widetilde{\Phi}(q,z) = A_j(q)e^{-ik_j(z-z_j)} + B_j(q)e^{ik_j(z-z_j)}, \quad z_j < z < z_{j-1},$$

$$A_{j+1}(q) = -\frac{1-r_{j,j+1}}{r_{j+1,j}r_{j+1}\exp(2ik_{i+1}d_{i+1}) - 1} \exp(ik_{i+1}d_{i+1}) A_j(q),$$

$$B_j(q) = -r_j A_j.$$

(S10)

Here $z_j = -\sum_{i=1}^{j} d_i$ is the $z$-coordinate of the bottom of $j^{\text{th}}$ layer. The recursion is initialized with $A_0(q) = \widetilde{\Phi}_0(q)$, where $\widetilde{\Phi}_0(q)$ is the Fourier transform of the scalar potential due to the incident field at the top surface $z = z_0 = 0$.

We used a similar set of equations to compute the field induced by the strip launcher discussed in the main text. We treated the launcher as a source located just inside the substrate. In this formulation, the potential created by the launcher is therefore incident from the bottom not the top of the structure. To adapt Eq. (S9) to this situation, the only change needed is switching the direction of the $z$-axis and renumbering of the layers in the opposite order (bottom to top). For the source potential, we used the known analytical solution for an ideal metallic strip of width $w$ subject to a unit uniform external field: $\Phi_0(x) = \text{Re}\left(x - \sqrt{x^2 - (w/2)^2}\right)$, $\widetilde{\Phi}_0(q) = i\pi a J_1(qw/2)/|q|$, where $J_1(z)$ is the Bessel function. Representative results for $E_z$ in the interior of the sample and on its top surface are shown in Fig. 2 of the main text and Fig. S6B, S6D, S6E and S6H-J below.

To model the response of h$^{11}$BN and MoO$_3$ we use the single Lorentzian oscillator form,

$$\varepsilon_X^i = \varepsilon_{\infty,X}^i \left(1 + \frac{\omega_{LO,X}^{i\,2} - \omega_{TO,X}^{i\,2}}{\omega_{TO,X}^{i\,2} - \omega^2 - i\omega\Gamma_X^i}\right), \quad i = x, y, \text{or } z,$$

(S11)

with parameters listed in Table S1.

Before discussing the bicrystal, it is instructive to review the case (*39*) of an $M = 2$ system - a single finite-thickness slab of an HM on a half-infinite substrate - for which a single recursion step of the calculation suffices. After some algebra, $r_p$ can be written as

$$r_p = -\frac{\sin\left[\frac{1}{2}\phi - \pi\alpha(\varepsilon_0)\right]}{\sin\left[\frac{1}{2}\phi + \pi\alpha(\varepsilon_0)\right]},$$

(S12)

$$\phi = q\delta + 2\pi\alpha(\varepsilon_s),$$

(S13)

$$\delta = 2d\tan\theta = 2id\frac{\sqrt{\varepsilon^x}}{\sqrt{\varepsilon^z}}.\tag{S14}$$

Here $\bar{\varepsilon}_s = \sqrt{\varepsilon_s^x}\sqrt{\varepsilon_s^z}$ is the effective permittivity of the substrate and $\varepsilon^x$, $\varepsilon^z$ are the in- and out-of-plane permittivities of the slab. The physical meaning of quantity $\phi$ is the phase accumulation of the polariton wave over a single "bounce" trajectory that involves traveling inside the slab from top to bottom, reflecting from the bottom surface, and returning to the top. The quantity $\delta$ is the lateral shift of the polariton rays over the same bounce. This important parameter will be discussed in more detail shortly.

As stated above, the waveguide mode momenta are the poles of function $r_p(q,\omega)$. As one can see from Eq. (S12), such poles arise whenever the sum $\phi + 2\pi\alpha(\varepsilon_0)$ approaches an integer multiple of $2\pi$. This is just the usual Bohr quantization rule that the total phase accumulation on a closed-cycle trajectory must be equal to $2\pi l$. There is an infinite number of such poles (*20, 40*):

$$q_l = \frac{2\pi}{\delta}[l - \alpha(\varepsilon_0) - \alpha(\varepsilon_s)].\tag{S15}$$

Admissible values of $l$ are determined from the condition $q_l = |q_x| \geq 0$. Note the sign of $\delta$ is the same as the sign of $\tan\theta$. Therefore, $l = 0, 1, 2, \ldots$ are allowed for a type-I HM with $\delta > 0$, such as our crystal A (h$^{11}$BN) and $l = -1, -2, -3, \ldots$ are allowed for a type-II HM (crystal B or MoO3), where $\delta < 0$. The first entries of these lists ($l = 0$ for type I and $l = -1$ for type II) are the principal modes. They are usually easiest to detect experimentally.

Taking the derivative in Eq. (S15), we find the in-plane group velocity $v \equiv v_x(q_x = q \geq 0)$ of the waveguide modes:

$$v = \frac{d\omega}{dq_l} = -\frac{\Omega(\omega)\tan\theta}{q_l}\left\{1 + \frac{\Omega(\omega)\tan\theta}{l - \alpha(\varepsilon_0) - \alpha(\varepsilon_s)}\frac{d}{d\omega}[\alpha(\varepsilon_0) + \alpha(\varepsilon_s)]\right\}^{-1}.\tag{S16}$$

The leading factor in this formula is identical to the first equation in Eq. (S5) for a bulk HM. The expression inside the braces represents a correction due to the reflection phase shifts $\pi\alpha(\varepsilon_0)$ and $\pi\alpha(\varepsilon_s)$ at the surfaces, i.e., the polaritonic Goos-Hänchen effect (*36*). Assuming this correction is not large, Eq. (S16) predicts that for a type-I HM (h$^{11}$BN) the sign of the in-plane component of the group velocity, $v$, of the waveguide modes is opposite in sign to the phase velocity. For a type-II HM (MoO3), the in-plane group and phase velocity components have the same sign.

Due to unavoidable dielectric losses characterized by the imaginary parts of the permittivity components $\varepsilon^i$, the momenta $q_l = q_l' + iq_l''$ and the velocities $v = v' + iv''$ given by Eqs. (S15), (S16) are in fact complex. The physically relevant real quantities are the mode wavelengths $\lambda_l = 2\pi/q_l'$ and the mode propagation lengths $L_l =$

$\operatorname{sgn} v'/q_l'' > 0$. Per Eq. (S15), both quantities scale approximately as $\lambda_l, L_l \propto 1/l$ with the mode index $l$. Hence, higher index modes have shorter propagation lengths. Distinguishing discrete eigenmodes remains possible if their momentum spacing $\Delta q = q_{l+1} - q_l = 2\pi/\delta$ exceeds their momentum broadening $q_l''$, which is equivalent to the condition that the propagation length $L_l$ is longer than $|\delta|$. This condition can be reinterpreted in terms of the ray picture. It means that the quantized waveguide modes exist only if polariton rays can survive multiple roundtrips between the two surfaces of the slab, see below.

Instead of working with real $\omega$ and complex $q$, we can restrict $q$ to be real, in which case the corresponding mode frequency $\omega_l(q) = \omega_l'(q) + i\omega_l''(q)$ must be complex, with some positive imaginary part $\omega_l'' > 0$, which plays the role of the mode linewidth. For weak losses, $\omega_l'' \simeq v' q_l''$. The mode quantization persists until $\omega_l''$, which is roughly $l$-independent, remains smaller than the intermode spectral gap

$$\Delta\omega \approx v'\Delta q \approx \frac{\pi}{qd}\Omega, \tag{S17}$$

which decreases with $q \propto l$.

To illustrate these properties on concrete examples we consider: i) crystal A (h$^{11}$BN) suspended in air and ii) crystal B (MoO$_3$) on SiO$_2$ substrate. These examples represent the top and bottom halves of our actual sample. The corresponding reflection coefficients are given by (cf. Eq. (S12))

$$r_p = -\frac{\sin\left[\frac{1}{2}\phi_X - \pi\alpha_X(\varepsilon_0)\right]}{\sin\left[\frac{1}{2}\phi_X + \pi\alpha_X(\varepsilon_0)\right]}, \quad X = A \text{ or } B, \tag{S18}$$

$$\phi_A \equiv q\delta_A + 2\pi\alpha(\varepsilon_0), \quad \phi_B \equiv q\delta_B + 2\pi\alpha(\varepsilon_s), \quad \delta_X \equiv 2d_X \tan\theta_X. \tag{S19}$$

The plots $\operatorname{Im} r_p$ for these subsystems are presented in Fig. S3A and S3B, respectively. The bright lines tracing the peaks of $\operatorname{Im} r_p$ are the dispersion curves as a function of $q > 0$. These curves have a negative dispersion for crystal A and a positive dispersion for crystal B, in agreement with the above determination.

In our discussion of a local source in a bulk HM we introduced two complementary concepts: modes and rays. The same can be done in the case of a finite-thickness slab. The ray picture is useful for understanding real-space field distributions at short distances from the source whereas the long-distance behavior is easier to analyze in terms of the waveguide modes. To see how the ray picture is modified due to the presence of the boundaries, we can expand the reflection coefficient $r_p$ in the infinite series

$$r_p(q,\omega) = -e^{2i\pi\alpha(\varepsilon_0)} - \left[1 - e^{2i\pi\alpha(\varepsilon_0)}\right] e^{2i\pi\alpha(\varepsilon_s)} \sum_{n=1}^{\infty} e^{2i\pi(n-1)[\alpha(\varepsilon_0)+\alpha(\varepsilon_s)]} e^{inq\delta}. \quad (S20)$$

(This series are converging because $\mathrm{Im}\,\delta > 0$ in the presence of losses.) Importantly, the only $q$-dependence of the coefficients is due to the factors $e^{inq\delta}$, which act as shift operators in the real space. This permits one to interpret the result in terms of the method of images: whereas in a bulk HM, a local source creates a field distribution peaked on a "resonant" conical surface, in the slab, the field is the superposition of this resonant cone with all its images obtained by successive reflections with respect to the top and bottom surfaces. The $x$-$z$ cross-section of this distribution consists of ray-like trajectories that bounce between the two surfaces maintaining the same angle $\pm\theta$ with respect to the $z$-axis. These hyperbolic rays return to each surface with regular intervals $\delta = 2d\tan\theta$, producing sharp peaks of the field.

As the rays gradually broaden with the distance travelled, the contributions of successive reflections start to overlap and an alternative description in terms of waveguide modes becomes more convenient. Indeed, as the distance from the source increases, high-$l$ modes, which have shorter propagation lengths, get rapidly damped, so that the long-distance pattern is dominated by the principal mode, a decaying sine wave. The period of this wave is $2\pi/q_0'$ (in type-I HM), which exceeds the ray repeat distance $|\delta|$ by a numerical factor. In most of near field imaging experiments, including ours, this regime is reached rather quickly, as shown in Fig. 2 of the main text and Fig. S6 below. Conversely, description of the short distance "ray behavior" in terms of waveguide modes is a bit less intuitive. It relies on the fact that the momentum spacing of the waveguide modes $\Delta q = q_{l+1} - q_l = 2\pi/\delta$ is equidistant, see Eq. (S15). Coherent beating of multiple such modes produces the sharp $\delta$-periodic peaks in the real space.

We are now ready to consider the case of a bicrystal, which we model as an $M = 3$ structure with layers $j = 1, 2,$ and $3$ representing crystal A (h$^{11}$BN), crystal B (MoO$_3$), and the substrate (SiO$_2$), respectively. The reflection coefficient $r_p$ of this system can be calculated using two recursion steps of Eq. (S7). After some algebraic manipulations, the result can be written as

$$r_p = -\frac{\sin\left[\frac{\phi_A + \phi_B}{2} - 2\pi\alpha_A(\varepsilon_0)\right] + r_{AB}\sin\left[\frac{\phi_A - \phi_B}{2} - 2\pi\alpha_A(\varepsilon_0)\right]}{\sin\left(\frac{\phi_A + \phi_B}{2}\right) - r_{AB}\sin\left(\frac{\phi_A - \phi_B}{2}\right)}, \quad (S21)$$

where $\phi_A$, $\phi_B$, and $r_{AB}$ are given by Eqs. (S8) and (S17). The plot of $\mathrm{Im}\,r_p$ calculated according to this formula is shown in Figs. S3C and S3D. In Fig. S3C, we use artificially reduced damping parameters to visualize the mode dispersion more clearly; in Fig. S3D, we use the realistic parameters from Table S1. The dispersion is complicated, consisting of numerous branches that are non-monotonic in $q$. As we discuss below, these branches result from hybridization and avoided crossing between the two families of waveguide modes seen in Figs. S3A and S3B.

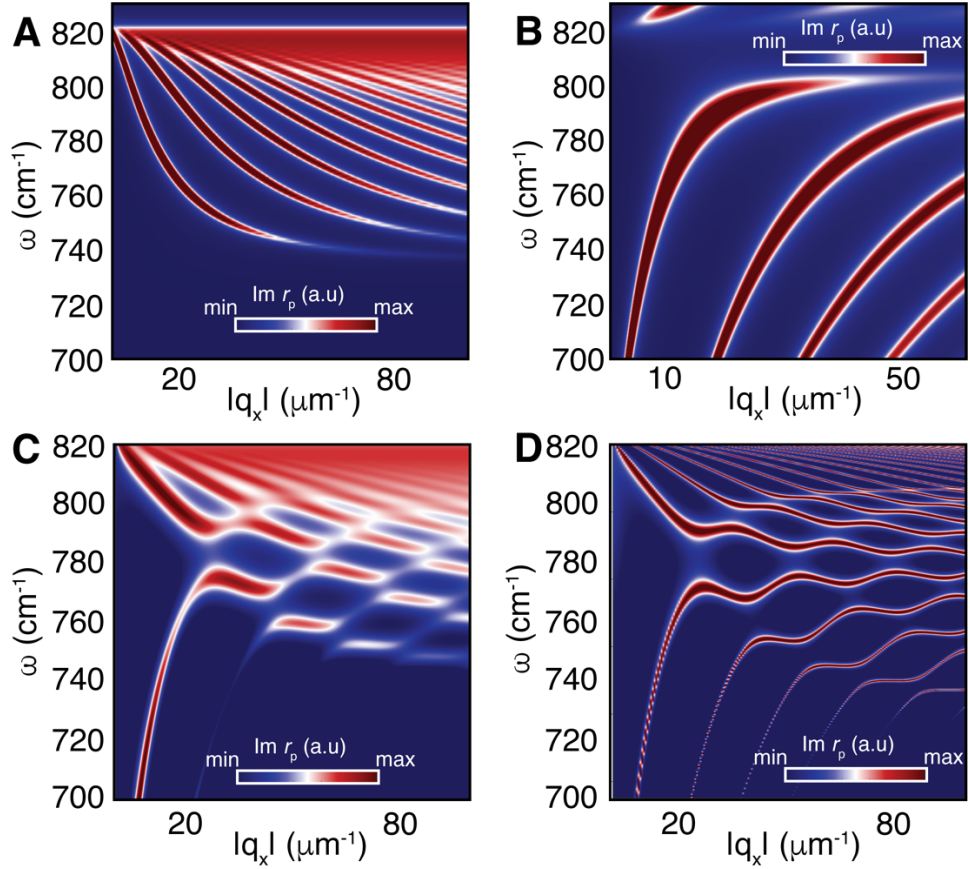

Figure S3 | **Waveguide mode dispersions in crystals A, B, and in the A-B bicrystal. A**, Mode dispersions of crystal A (h¹¹BN). **B** Same for crystal B (MoO$_3$). **C**, and **D**, Dispersions in the A-B bicrystal with realistic and reduced losses, respectively. The loss parameters $\Gamma$ are artificially reduced by a factor of $10^3$ compared to Table S1.

According to Eq. (S21), the poles of $r_p$, which are the mode dispersions, are the roots of the following equation:

$$\sin\left(\frac{\phi_A + \phi_B}{2}\right) - r_{AB} \sin\left(\frac{\phi_A - \phi_B}{2}\right) = 0 . \tag{S22}$$

To analyze mathematical properties of these roots we treat the A-B interface reflection coefficient $-1 \leq r_{AB} \leq 1$ as a fixed real parameter and solve Eq. (S20) for $\phi_A$ and $\phi_B$. A few of such solutions are plotted in Fig. S4, using the sum $\phi_+ \equiv \phi_A + \phi_B$ and difference $\phi_- \equiv \phi_A - \phi_B$ phase variables as the coordinates on the axes. For each $r_{AB}$, the solution consists of an infinite number of curves of the same shape which are related by periodic translations. In three cases $r_{AB} = -1, 0, 1$, the curves straighten into lines, e.g., for the reflectionless interface $r_{AB} = 0$ these are the horizontal lines $\phi_+ = 2\pi m$, where $m$ is an integer. When $r_{AB}$ deviates from zero, a special set of points on these lines, where $\phi_+$ and $\phi_-$ are both integer multiples of $2\pi$ remain solutions of Eq. (S22). In between such "anchor" points, the curves move either closer or further away from their neighbors along the vertical ($\phi_+$) direction. The minimal distance in $\phi_+$ between the nearest neighbor curves is

$$\Delta\phi_+ = 4\arccos |r_{AB}|, \tag{S23}$$

which is a decreasing function of the A-B interface reflectivity. If $|r_{AB}| = 1$, then $\Delta\phi_+$ vanishes; if $r_{AB} = 0$, then $\Delta\phi_+$ has the maximum possible value of $2\pi$.

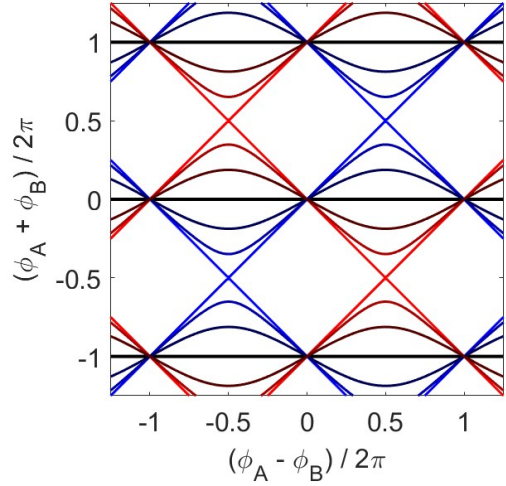

Figure S4 | **Waveguide mode dispersion of a bicrystal, expressed in phase variables.** The line colors mark the A-B reflection coefficient $r_{AB}$ that varies from $-1$ (blue) to $+1$ (red) in six increments. The reflectionless interface $r_{AB} = 0$ (black horizontal lines) produces the largest possible spectral gaps.

The interpretation of the described structure in terms of eigenmode coupling is straightforward. If $|r_{AB}| = 1$, then the interface is impenetrable, and so the crystal A and B are isolated from one another. Their mode dispersion lines (the diagonal lines in Fig. S4) can freely intersect, which means that the dispersion of the whole bicrystal is gapless. If $|r_{AB}| < 1$, the A-B interface is partially transparent, so that the two subsystems interact. This interaction causes spectral repulsion and the gaps in the dispersion. If the interface is fully transmitting, $r_{AB} = 0$, then the mode coupling is the strongest and the gaps are the largest. In this latter case the two subsystems are perfectly impedance-matched, so that they behave as a single slab with $\delta$ equal to $\delta_A + \delta_B$. Indeed, using Eq. (S19), it is easy to check that the mode quantization condition $\phi_+ = 2\pi m$ that applies for $r_{AB} = 0$ is equivalent to the following modification of Eq. (S15):

$$q = \frac{2\pi}{\delta_A + \delta_B}[m - \alpha(\varepsilon_0) - \alpha(\varepsilon_s)] \quad (r_{AB} = 0). \tag{S24}$$

Here we dropped the subscripts of $\alpha$'s because $\alpha_A = \alpha_B$ in this case.

Turning to the group velocity component $v$, for the (nearly) impenetrable interface case $|r_{AB}| \simeq 1$, it alternates as a function of $q$ between two values, $v_A$ and $v_B$, given by Eq. (S15):

$$v_X = -\frac{\Omega_X(\omega)\tan\theta_X}{q}, \qquad X = A \text{ or } B \qquad (S25)$$

(for simplicity, we neglected the Goos-Hänchen correction). Since $v_A < 0$ and $v_B > 0$, the group velocity changes sign at each anti-crossing. For the reflectionless case, the group velocity of each branch does not alternate in sign. It is given by the expression

$$v = \frac{\delta_A + \delta_B}{\delta_A v_A^{-1} + \delta_B v_B^{-1}} \qquad (r_{AB} = 0) \qquad (S26)$$

whose denominator is always negative. Hence, the sign of $v$ is opposite to that of the numerator $\delta_{tot} = \delta_A + \delta_B$. The plot of $\delta_{tot}(\omega)$ computed for our experimental system (Fig. S5) indicates that it is an increasing function that crosses zero at frequency $\omega_0 = 785$ cm$^{-1}$.

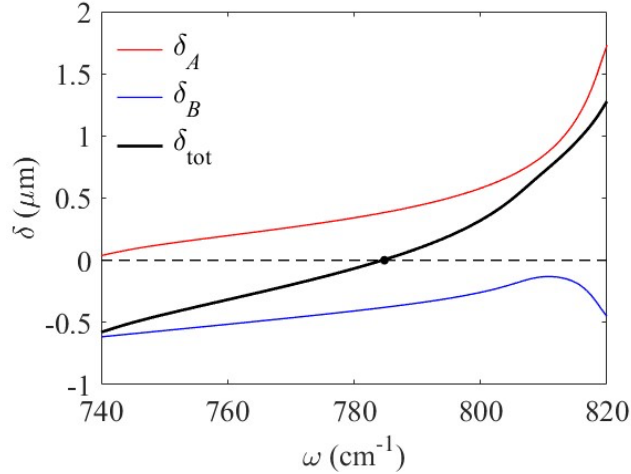

Figure S5 | **Calculated ray displacements (real parts) in the bicrystal** (see text).

In general, $\omega_0$ is the solution of the equation

$$\delta_{tot}(\omega_0) = \delta_A(\omega_0) + \delta_B(\omega_0) = \frac{\sqrt{\varepsilon_A^x(\omega)}}{\sqrt{\varepsilon_A^z(\omega)}}d_A + \frac{\sqrt{\varepsilon_B^x(\omega)}}{\sqrt{\varepsilon_B^z(\omega)}}d_B = 0, \qquad (S27)$$

so it depends on the thickness ratio $d_A/d_B$ of the two crystals. Frequency $\omega_0$ plays the role of the separatrix dividing the regions of predominantly positive and predominantly negative dispersion. Frequency $\omega_0$ also has an intuitive interpretation within the semiclassical ray picture. At frequency $\omega_0$ the ray orbits become closed (or periodic), see Fig. 1C, suggesting that $v$ should be zero, in agreement with Eq. (S26). Finally, we give the formula for the spectral gaps separating the dispersion curves:

$$\Delta\omega = v\Delta q = \left(\frac{1}{\Delta\omega_A} + \frac{1}{\Delta\omega_B}\right)^{-1} \propto \frac{1}{q} \qquad (r_{AB} = 0). \tag{S28}$$

Here the gaps $\Delta\omega_X$ of each subsystem in isolation are given by Eq. (S17).

In the actual bicrystal, $|r_{AB}|$ varies with frequency and is typically somewhere in between 0 and 1. Therefore, the numerically calculated dispersions seen in Fig. S3C and S3D exhibit behavior intermediate between the two limits described above. For example, the coupling of the principal modes of Crystal A and B produces the dispersion lines with sign alternating slopes, as appropriate for a non-negligible $r_{AB}$. Combining Eqs. (S23) and (S28), we estimate the size of the gap for these principal modes to be

$$\Delta\omega \sim \frac{2}{\pi}\left(\frac{1}{\Delta\omega_A} + \frac{1}{\Delta\omega_B}\right)^{-1} \arccos |r_{AB}|. \tag{S29}$$

Strong coupling of the eigenmodes is realized, by definition, when the mode hybridization gap exceeds their combined linewidth, i.e., if $\Delta\omega > \omega_A'' + \omega_B''$. Since $\Delta\omega$ decreases with $q$ while $\omega_X''$ stays roughly constant, the latter condition is more difficult to achieve for higher-order modes. Using the parameters from Table S1, we conclude that this condition is satisfied for the crossing of the principal modes only.

*S1.3 Numerical simulations of layered heterostructures*

To model the real-space fringe pattern observed via s-SNOM near a $MoO_3$ edge, we assumed that the polaritons were launched by the sharp $MoO_3$ edge alone, with the s-SNOM tip acting only as a detector. We further assumed that the measured complex near-field signal $se^{i\theta}$ is proportional to the out-of-plane field component $E_z$. To simplify the calculation of the field component, $E_z$, we neglected variation of all the quantities along the $y$-direction, parallel to the edge of the $MoO_3$. As in Eq. (S9), we adopted the quasistatic approximation $E_z = -\frac{\partial}{\partial z}\Phi(x,z)$, where $\Phi(x,z)$ is the scalar potential obeying the equation

$$\left[\frac{\partial}{\partial x}\varepsilon^x(x,z)\frac{\partial}{\partial x} + \frac{\partial}{\partial z}\varepsilon^z(x,z)\frac{\partial}{\partial z}\right]\Phi(x,z) = 0. \tag{S30}$$

The solution of Eq. (S30) was computed numerically using the MATLAB PDE Toolbox. We took the simulation domain to be a $4 \times 1.3 \ \mu m^2$ rectangle subdivided into layers of different materials as depicted in Fig. S6. To include a uniform external electric field $\vec{E}_0$ incident at angle $\pi/4$ in the $x$-$z$ plane we used the boundary condition $\Phi = -\vec{E}_0 \cdot \vec{r}$ at the edges of the domain. The frequency dependence of the solution comes from that of the permittivity components $\varepsilon^x(x,z)$ and $\varepsilon^z(x,z)$.

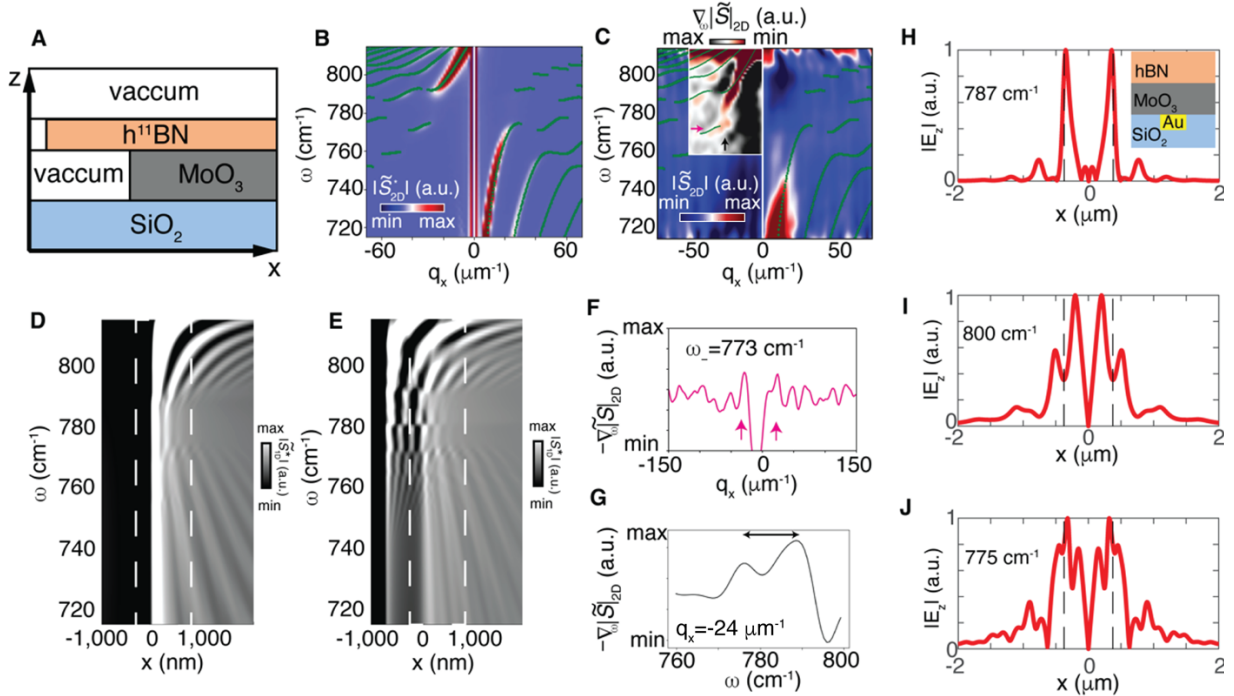

Figure S6 | **Numerical solutions of Eq. (S30) and comparison with experimental data. A**, Geometry used to model the experiments with edge-launched polaritons. **B**, Amplitude of a two-dimensional Fourier transform, $|\tilde{S}_{2D}^*|$, of the calculated complex electric field component, $E_z$, evaluated at the top surface of the bicrystal. The thicknesses in the calculations are 58 nm (h[11]BN) and 150 nm (MoO$_3$), appropriate for Fig. 3 of the main text and panel (**C**) in this figure. **C**, The amplitude of the two-dimensional Fourier transform of nano-optical data $|\tilde{S}_{2D}(q_x, \omega)|$ is shown with a red-white-blue colormap. The inset shows the frequency derivative $\nabla_\omega |\tilde{S}_{2D}(q_x, \omega)|$ with a red-white-gray colormap, to sharpen subtle features in the data. These data were obtained on the same crystal studied in Fig. 3 of the main text. The green dots in panels (**B**) and (**C**) show locations where maxima are observed in calculations of Im $r_p$ (see text). **D**, and **E**, Amplitude of a one-dimensional Fourier transform, $|\tilde{S}_{1D}^*|$ of the calculated electric field component, $E_z$. **D**, Both the h[11]BN and MoO$_3$ edges are at $x = 0$. **E**, The h[11]BN edge is located at $x = -700\ nm$, while the MoO$_3$ edge is located at $x = 0$, to replicate the conditions in our experiments (see Fig. S9). The white dashed lines are used to indicate the region where experimental data are shown in Fig. 3B of the main text. **F**, A linecut of $-\nabla_\omega |\tilde{S}_{2D}(q_x, \omega)|$ taken at constant frequency $\omega_- = 773 +/-3\ cm^{-1}$ (magenta arrow in panel **C**). The arrows mark the positions where peaks are anticipated from our calculations. **G**, Linecut taken at a constant momentum $q_x = 24\ \mu m^{-1}$ (black arrow in panel **C**) shows extrema of $-\nabla_\omega |\tilde{S}_{2D}(q_x, \omega)|$, split by $\Delta\omega = 13 +/-3\ cm^{-1}$ as we indicate with a black arrow. **H-J**, Calculated profiles of the normal electric field component produced by strip-launched polaritons just above the hetero-bicrystal at frequencies 787, 800, and 775 $cm^{-1}$, respectively. The crystal thicknesses in the calculations are 98 nm (h[11]BN) and 290 nm (MoO$_3$), appropriate for Fig. 2 of the main text and Fig. S7 and S8 below.

We proceed to discuss simulations of the edge launched experiments and the comparison with our experimental data. We solved Eq. (S30) repeatedly on a grid of frequency values and the results are presented in Fig. S6. We stress that owing to the 700 nm lateral displacement between the h[11]BN and MoO$_3$ edges, interference between bicrystal polaritons and polaritons launched by the natural h[11]BN edge slightly modifies their dispersion (see Fig. S6D and S6E). The dashed lines of Fig. 3A of the main text show the locations where maxima are observed in the calculations of Fig. S6D, obtained for a simplified model where h[11]BN and MoO$_3$ edges are aligned. The simplified model captures the salient features of the experimental data. The two-dimensional Fourier transform in Fig. S6B was calculated with the 700 nm displacement and shows excellent agreement with the experimental dispersion in Fig. S6C. The dispersion relationship obtained from the maxima in calculations of Im $r_p$ (green dots) is superimposed on our data and calculations to highlight this agreement. In addition to the dispersion of the principal branch, reported in Fig. 3 of the main text, features associated with higher order modes are also identified in 'hBN-like' portion of the dispersion in Fig. S6B and S6C. We stress that the dispersion of Im $r_p$ are calculated as a function of the absolute value of the momentum. The results are plotted for negative or positive $q_x$ if the dispersion is negative or positive respectively.

The electric field profiles calculated for the strip-launched polaritons are presented in Fig. S6H-J. For simplicity, these calculations were performed using the semi-analytic recursion procedure (Eq. (S9)) instead of the fully numerical PDE solver. The calculations reproduce the main qualitative features of the data: (1) rapid decay of the polaritonic oscillations away from the launcher at $\omega_0 = 787$ cm$^{-1}$ vs. their gradual decay above and below this frequency (2) lone vs. twin peaks of electric field above the strip edges, $x = \pm w/2$, at and away from $\omega_0$, respectively (3) enhanced electric field above the Au strip, $-w/2 < x < w/2$, compared to that above the un-patterned SiO$_2$ substrate. We note however one key discrepancy between the numerical simulations and the experimental data. By symmetry, the solution of Eq. (S30) for $E_z(x)$ is an odd function vanishing at $x = 0$. However, a non-zero SNOM signal is measured at $x = 0$ in the experiments (Fig. 2 of the main text). One of the possible reasons for this discrepancy is that our model does not include the tip-launched waves. The tip contribution to $E_z(x)$ is even, so that the total $E_z(0)$ can indeed be nonvanishing. Qualitatively, the tip-induced field should display a contrast between Au-patterned and un-patterned parts of the substrate and, in addition, $\lambda_p/2$-periodic fringes caused by reflections of tip-launched polaritons off the strip edges. A quantitatively accurate calculation of this field profile remains a challenge for the theory.

Our numerical simulations and experimental Nano-FTIR data are compared in Fig. S6C. These data are simply another representation of the previously displayed hyperspectral data of $|\tilde{S}_{1D}(X, \omega)|$ shown in Fig. 3B of the main text. Here, we adopted a two-dimensional Fourier transform analysis of Nano-FTIR data, as introduced in Ref. (*27*). The analysis involves performing the Fourier transform of $\tilde{S}_{1D}(X, \omega)$ along the $x$-axis and taking its absolute value, which yields $|\tilde{S}_{2D}(q_x, \omega)|$. The maxima of $|\tilde{S}_{2D}(q_x, \omega)|$ reveal the polariton dispersion $q_x = q_x(\omega)$ of the principal mode, including both the magnitude and sign of $q_x$. By considering $\nabla_\omega |\tilde{S}_{2D}(q_x, \omega)|$, in the inset of Fig. S6C subtle polaritonic features are sharpened over the background noise. Note that $\nabla_\omega |\tilde{S}_{2D}(q_x, \omega)|$ displays inflection points at the frequencies where $|\tilde{S}_{2D}(q_x, \omega)|$ is maximized. Thus, the $(q_x, \omega)$ values corresponding to the principal mode are slightly red shifted from the locations where minima are observed in $\nabla_\omega |\tilde{S}_{2D}(q_x, \omega)|$ in Fig. S6C. Good agreement between the experimental data and numerical calculations is readily observed. This includes a few data points associated with higher-order branches identified in the 'hBN-like' portion of the dispersion in the inset.

The data in Fig. S6C establish that a change in sign of $q_x$ is associated with the crossover between positive and negative dispersion in the bicrystal. The negative dispersion on the left-hand side of the plot in Fig. S6C derives from h[11]BN whereas positive dispersion on the righthand side stems from MoO$_3$. Polaritons with 'h[11]BN-like' character ($q_x < 0$) display a gap in their dispersion at $\omega_0$ (Fig. 3; Fig. S6B, S6C, S6G) and, notably, persist at a slightly lower frequency, $\omega_-$ (Fig. 3; Fig. S6B, S6C, S6F). 'MoO$_3$-like' modes ($q_x > 0$), which are also observed at $\omega_-$, abruptly vanish at $\omega \geq \omega_0$. To highlight these features, we show linecuts from the data in Fig. S6C in Figs. S6F and S6G. First, in Fig. S6F we show a linecut of $\nabla_\omega |\tilde{S}_{2D}(q_x, \omega = \omega_- = 773 \text{ cm}^{-1})|$. The extremum identified near $q_x = 26 \ \mu\text{m}^{-1}$, marks the MoO$_3$-like polariton. A second extremum is also detected near $q_x = -28 \ \mu\text{m}^{-1}$. Thus, at least two modes are detected at $\omega_-$, one with positive and one with negative $q_x$. A linecut of $\nabla_\omega |\tilde{S}_{2D}(q_x = -24 \ \mu\text{m}^{-1}, \omega)|$ reveals a two-peak profile, indicating a gap in the polariton dispersion. The $\Delta\omega \cong 13 \text{ cm}^{-1}$ magnitude of the gap, reported in the main text, is indicated with a black arrow in Fig. S6G. The error quoted in the main text represents the spectral resolution of the measurement. Our observations establish mode repulsion at $\omega_0$ and waves with both positive and negative dispersion at $\omega_-$.

## S2. Extended data
*S2.1 Near-field experiments with strip-launched polaritons*

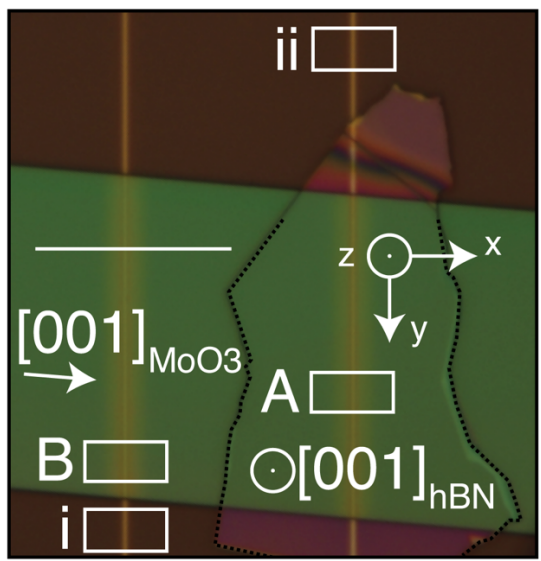

Figure S7 | **Microscope image of the device used in experiments with Au launchers**. The white scale bar is about 34 μm in length (See text for details).

Figure S7 shows a microscope image of the device studied in our experiments where polaritons were launched by Au strips, Fig. 2 of the main text. The strips were patterned on an SiO$_2$ substrate. Next, crystal B (MoO$_3$) was transferred on top of the Au launchers, covering them in both regions i and ii, with its c-axis nearly perpendicular to the facets of the strips. Finally, crystal A (h$^{11}$BN) was transferred on top of crystal B (MoO$_3$) forming the h$^{11}$BN /MoO$_3$ bicrystal. Images collected inside regions roughly indicated by the white boxes A and B are shown in Fig. 2 of the main text. The data shown in Fig. S8 were acquired on the same device, with panels A-C obtained within region B and panels D-F within region A.

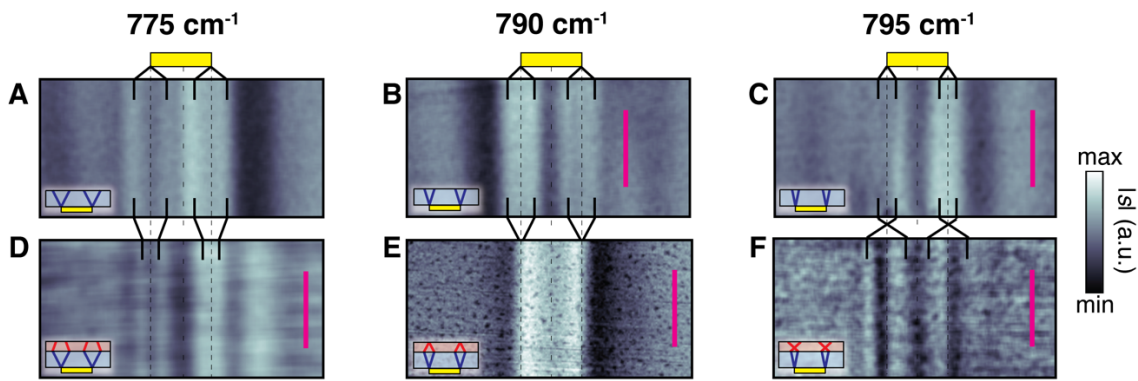

Figure S8 | **Near-field images and ray trajectories.** images collected on the device shown in Fig. S7 at ambient temperature. **A-C** images collected at the MoO$_3$ surface. **D-F** images obtained at the h$^{11}$BN/MoO$_3$ surface. The black lines are guides to the eye that mark the maxima observed in the images

of the near-field amplitude, |s|. Illustrations of the inferred ray trajectories are sketched in the insets. The frequency of incident radiation and the position of the underlying Au launcher are indicated at the top of each column. Scale bars are indicated with magenta lines in each panel, which are 1 μm in length. The schematic insets depict the geometry in the xz-plane.

Figure S8 shows data collected with the bicrystal at three infrared frequencies: below, at, and above $\omega_0 \cong 790 \text{ cm}^{-1}$. To remove (extrinsic) slowly varying backgrounds the data shown in Fig. S8 were Fourier filtered for clarity, whereas unfiltered data are shown in Fig. 2 of the main text. The top and bottom rows correspond to the top surfaces of MoO$_3$ alone and h$^{11}$BN /MoO$_3$, respectively. Data obtained at a frequency $\omega = 775 \text{ cm}^{-1}$, which is below $\omega_0$, are shown in Figs. S8A and S8D; data obtained at frequency $\omega = 795 \text{ cm}^{-1}$ which is above $\omega_0$. Images obtained at $\omega_0$ are shown in Figs. S8C and S8F. Note that the data in Fig.S8 were obtained at room temperature.

We identify the maxima of $|s|$, indicated by the black lines used as guides for the eye, with the locations where polariton rays launched by an underlying Au strip reach the top surface. In MoO$_3$ polaritonic rays are detected at a finite lateral distance away from the underlying gold strip. This distance decreases as the infrared frequency increases, consistent with the dispersion of MoO$_3$ in Fig. S11. At the bicrystal surface, the displacement of polaritonic rays relative to the edge of the strip is non-monotonic with frequency. Specifically, at frequencies below $\omega_0$ the rays are observed away from the edges of the underlying gold strip. At $\omega_0$ the maxima of $|s|$ are nearly re-aligned with the lateral positions of the underlying edges of the gold launcher. Above $\omega_0$ the maxima of $|s|$ are, again, displaced from the edges of the underlying strip. The associated polaritonic ray trajectories are consistent with negative refraction as illustrated in the schematic insets.

*S2.2 Near-field experiments with edge-launched polaritons*

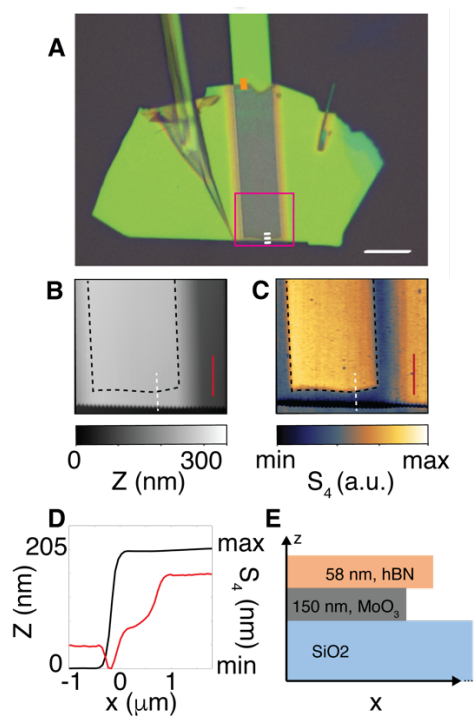

Figure S9| **Experiments with edge-launched polaritons. A**, Microscope image of the device. The magenta rectangle indicates the region where the data in panels (**B**) and (**C**) were recorded. **B**, AFM topography image showing the height, Z, of the sample. **C**, White-light near-field image showing the fourth harmonic of the scattering amplitude, $S_4$, collected with a broadband light source (see Fig. S14) at ambient temperature. The perimeter of $MoO_3$ is outlined with a black dashed line in panels (**B**) and (**C**). **D**, Co-located line profiles of Z and $S_4$ taken along the white dashed lines indicated in panels (**B**) and (**C**). **E**, Schematic illustrating the cross-section of the sample along the vertical plane that passes through the white dashed lines indicated in panels (**B**) and (**C**).

Bicrystal dispersion data in Fig. 3 of the main text were obtained on the sample shown in Fig. S9. Optical contrast associated with the $SiO_2$ substrate, a rectangular $MoO_3$ crystal, an overlying thin $h^{11}BN$ crystal and the $h^{11}BN/MoO_3$ bicrystal are identified in the optical microscope image shown in Fig. S9A. A topographic scan recorded near the bicrystal edge, within the magenta rectangle in Fig. S9A, is shown in Fig. S9B. At the bottom of the image, the substrate establishes a baseline for the topography (Z=0). An increase of topographic height marks the sample's edge. While gradual changes are identified in topography data collected within the interior of the sample surrounding the boundaries of $MoO_3$, sharp topographic features marking this boundary are not clearly observed. On the other hand, the optical data in Fig.S9C, recorded in the same region, reveals clear optical contrast between regions where $MoO_3$ is present and surrounding areas where it is not. The optical contrast image in Fig.S9C was therefore, used to determine boundaries of $MoO_3$. These boundaries are indicated with black dashed lines in Fig. S9B and S9C. Co-located linecuts of the data in Fig. S9B and S9C are shown in Fig.S9D. The optical contrast, associated with the edge of $MoO_3$, lags the single-step edge identified in topography by a lateral distance of about 700 nm. These data are all consistent with the schematic in Fig. S9E. The thin h$^{11}$BN crystal, on top of $MoO_3$, overhangs the underlying $MoO_3$ crystal edge for a small lateral distance. The sharp edge observed in topography marks the edge of h$^{11}$BN while the boundary of $MoO_3$, which we refer to as the 'bicrystal edge' in the main text and in Fig. S6, is located about 700 nm away from the h$^{11}$BN edge. We stress that this interpretation is consistent with the entire image of Fig. S9C. On the right-hand side of the image only h$^{11}$BN is present. The measured change in the topographic height on the right-hand side of Fig.S9B (~60 nm) is consistent with the 58 nm thickness of h$^{11}$BN measured at the boundary of h$^{11}$BN well away from the bicrystal and along the h$^{11}$BN/MoO3 boundary indicated with the orange line in panel A. Near the center of the topographic image in Fig. S9B (white dashed line) we witness a sharp change in the topographic height, with a larger magnitude of about 201 nm, at the sample/substrate interface. The height gradually increases to 208 nm in the interior of the bicrystal, consistent with the total combined thickness of 58 nm h$^{11}$BN and 150 nm $MoO_3$, which were independently measured with topographic scans well away from the bicrystal edge on the same sample.

The specific linecut where the hyperspectral data were acquired is indicated with the white dashed line. Here the $MoO_3$ and h$^{11}$BN edges are nearly parallel. Further, the [001]$_{MoO3}$ direction is nearly perpendicular to the h$^{11}$BN edge. Finally, we note that the edges of h$^{11}$BN and $MoO_3$ are nearly aligned. We note that '$MoO_3$-like' polaritons with positive dispersion were not observed in experimental hyperspectral dispersion data shown in Fig. S10A, which were obtained along the edge of only h$^{11}$BN shown with the orange line in Fig. S9A. To explain these results, we performed numerical simulations with the same geometry as in the experiment, shown in Fig. S10B. The numerical results confirm the qualitative aspect of our observations, that 'h$^{11}$BN-like' modes are launched with higher intensity than '$MoO_3$-like' modes. These results are readily rationalized by considering that h$^{11}$BN and $MoO_3$ are nearly impedance matched within the overlapping Reststrahlen band. Thus, the h$^{11}$BN/$MoO_3$ edge is an inefficient polariton launcher. It is possible that '$MoO_3$-like' modes could be detected at this edge in more careful experiments.

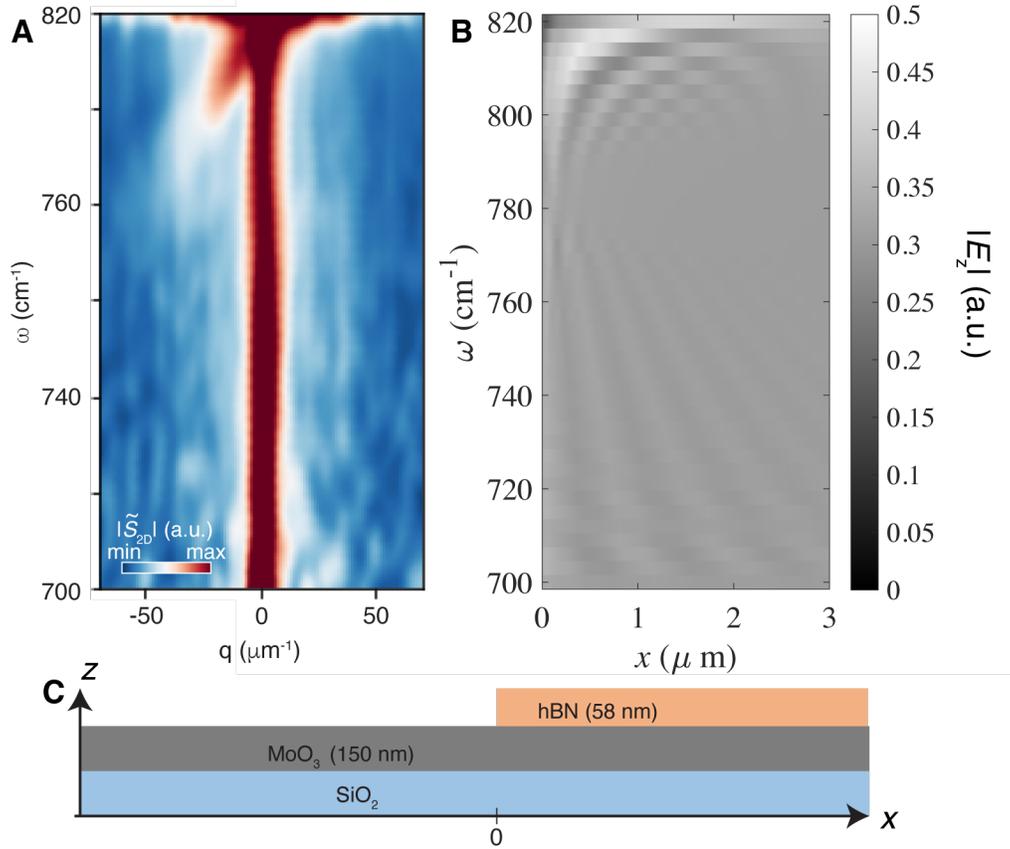

Figure S10| **Polariton dispersions at the h¹¹BN edge. A**, The amplitude of the two-dimensional Fourier transform of nano-optical data obtained at the bicrystal surface $|\tilde{S}_{2D}(q_x,\omega)|$ is shown as a function of the momentum component $q_x$ and frequency of the incident light, $\omega$. These data were recorded along the orange line in Fig.S9A. Along this line the h¹¹BN has an edge, while the underlying MoO₃ crystal runs continuously. **B**, Numerical calculations of the z-component of the electric field, $|E_z|$, which are solutions to Eq. (S30). **C**, Geometry of the device in our experiment, panel (**A**), and calculations, panel (**B**).

We also measured the dispersions of the individual h¹¹BN and MoO₃ crystals. These data can be represented as a hyperspectral map of $|\tilde{S}_{1D}(x,\omega)|$ where oscilations normal to the crystal edge reveal polaritons. Taking the absolute value after an additional Fourier transform of the complex quantity, $\tilde{S}_{1D}(x,\omega)$, is taken along the $x$-axis yields $|\tilde{S}_{2D}(q_x,\omega)|$. The maxima of this quantity correspond to the polariton momenta. We refined the optical constants of our crystals (Table S1 and Fig. 1A of the main text) to obtain a good agreement between observed and calculated dispersions. The far-field reflectance data of Fig. S13 was another input into this fitting process. With these parameters, we calculated the full dispersion relationship of the hetero-bicrystal polaritons shown in Fig. 3.

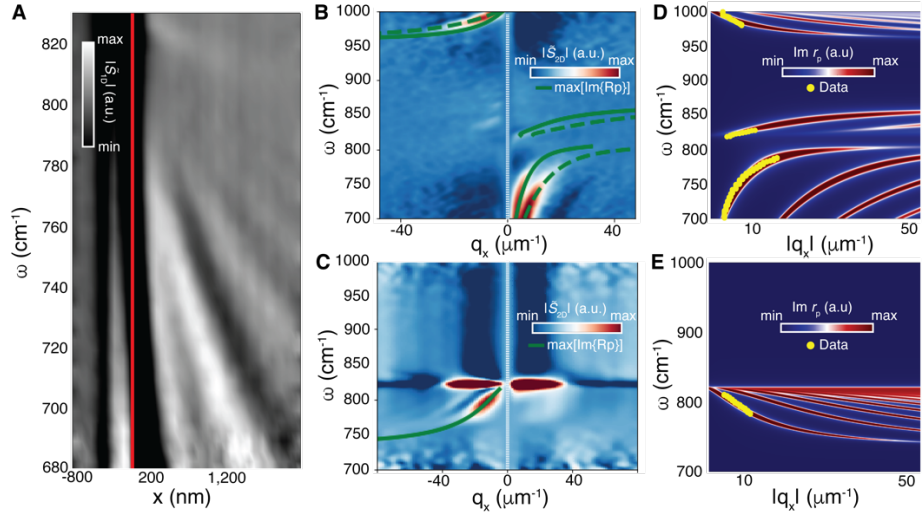

Figure S11| **Polariton dispersions of MoO₃ and h¹¹BN crystals. A**, The amplitude of the one-dimensional Fourier transform of nano-optical data obtained at the surface of MoO₃ $|\tilde{S}_{1D}(q_x,\omega)|$ is shown as a function of the momentum component $q_x$ and frequency of the incident light, $\omega$. The red solid line marks the crystal's physical edge. **B**, and **C**, The amplitude of the two-dimensional Fourier transform of nano-optical data $|\tilde{S}_{2D}(q_x,\omega)|$, obtained at the surface of **B**, MoO₃ with 150 nm thickness and **C**, h¹¹BN with 98 nm thickness. The white lines show $q_x$=0. The green lines show locations where maxima are observed in calculations of Im $r_p$ for $\lambda$ (solid lines) and $\lambda/2$ modes (dashed lines). **D**, and **E**, Calculations of Im $r_p$ as a function of $\omega$ and the absolute value of the momentum, $|q_x|$. **D**, for MoO₃, experimental data from panel (**B**) are shown with yellow dots. **E**, for h¹¹BN, experimental data from panel (**C**) are shown with yellow dots.

*S2.3 Experiments with disk-launched polaritons*

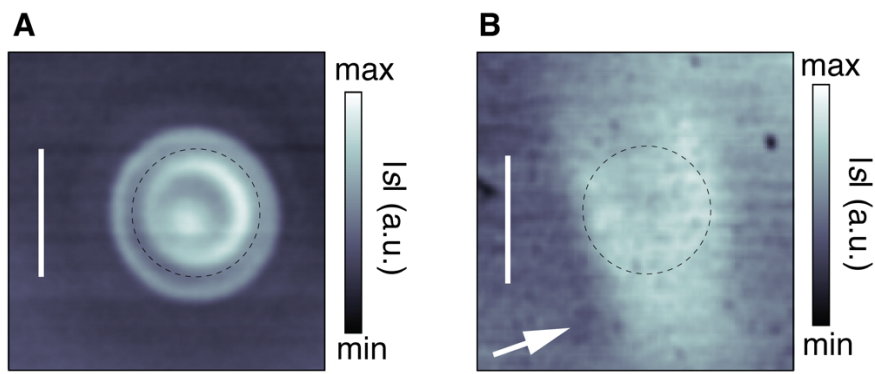

Figure S12| **Experiments with disk launched polaritons. A**, Experimental near-field amplitude data |s| at the surface of h$^{11}$BN placed on top of an Au disk launcher with a diameter of about 1 μm. **B**, |s| measured at the surface of the h$^{11}$BN/MoO$_3$/Au disk assembly. The c-axis of MoO$_3$ is indicated with a white arrow. The white scale bars in (**A**) and (**B**) are 1 μm in length. The data in both (**A**) and (**B**) were collected at ω=777+/-4 cm$^{-1}$ at ambient temperature. The approximate boundary of the disks is indicated in (**A**) and (**B**) with black dashed lines.

In this section we present two-dimensional images of polaritons excited by a disk-shaped launcher. Near-field images obtained on the top surface of h$^{11}$BN reveal 'hot rings' surrounding the edges of the Au disk (*40*). Their concentric shape is consistent with in-plane symmetry of the material. MoO$_3$ is orthorhombic. At 775 cm$^{-1}$ only the [001] component of the dielectric tensor is negative and the crystal is, therefore, in-plane hyperbolic (*38*). In Fig. S12B we show an image obtained on the top surface of an h$^{11}$BN/MoO$_3$ hetero-bicrystal. The image in Fig. S12B demonstrates that polaritons launched by the underlying disk propagate along the [001] direction of MoO$_3$. No polaritons propagating along the orthogonal [100] axis have been detected. These observations are consistent with the notion that the hetero-bicrystal inherits its in-plane hyperbolicity from MoO$_3$.

    Our observations in Fig. S12 suggest that it is possible to reduce the symmetries of propagating polaritons in hetero-bicrystals. Moreover, atomically layered structures are amenable to various forms of tuning once these structures are augmented by plasmonic (*39*), or photo-susceptible (*41, 42*), layers. These structures could, potentially, also be monolithically integrated with Si photonics and electronics (*43*) in future works.

*S2.4 Far field reflectivity of h¹¹BN and MoO₃*

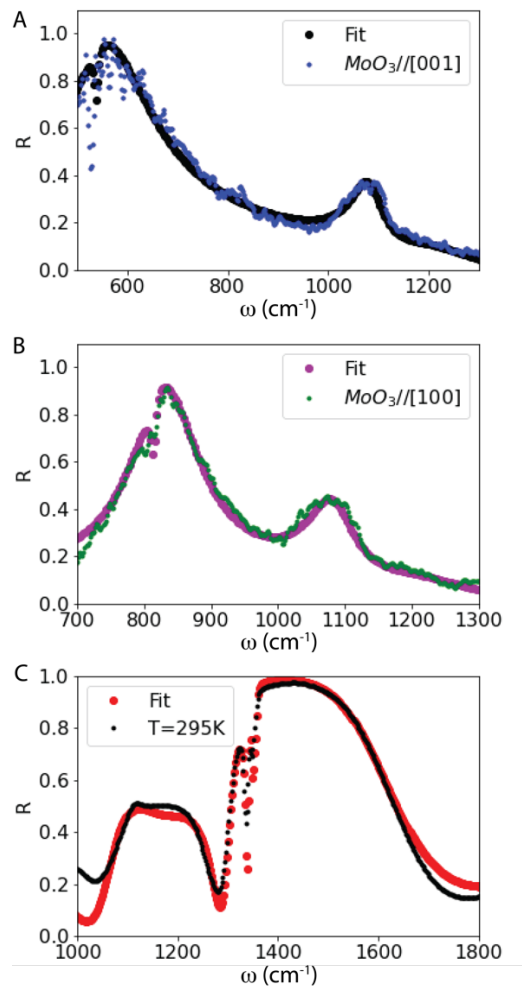

Figure S13| **Far-field Reflectivity spectra of h¹¹BN and MoO₃. A-C** Far-field reflectivity of h¹¹BN and MoO₃. **A-B**, Reflectivity spectra on MoO₃. **A**, blue dots show data obtained with the polarization along the [001] axis. A fit with parameters in Table S1 is shown with black points. **B**, Green dots show data obtained with the polarization along the [100] axis. A fit with parameters in Table S1 is shown with magenta points. **C**, Black dots show the reflectivity spectra obtained on h¹¹BN. A fit with parameters in Table S1 is shown with red points.

The near-field data in the main-text were augmented with far-field reflectivity measurements to determine the optical constants in Table S1. It can readily be appreciated that measurements of the dispersion relationship of phonon-polaritons, such as those in Fig. 11, are insufficient to determine the full dielectric tensor since the wavevector of propagating polaritons depends on both the in and out of plane components of the permittivity at a particular frequency. Infrared reflectivity provides a second measurement, which can be used to determine the optical constants within the xy-plane of the experiment, along the axis of polarization of the reflected light. The pair of measurements allows us to determine the two unknown quantities, namely the in- and out-of-plane components of the dielectric tensor, at each frequency.

*S2.5. Ultranarrow band mid-infrared beamline from a pulsed light source*

Here we describe the procedure used to generate the ultra-narrowband light used in imaging experiments. The spectral intensity of various sources used in this work are shown in Fig. S14. First, we show the spectra from our broadband source, generated by combining a tunable idler beam with a bandwidth around 250 cm$^{-1}$ with a 100 cm$^{-1}$ broad 1600 nm pump beam in a 500-micron thick GaSe crystal, in black in Fig. S14. The broadband radiation has full width at half maximum spectral width of around 250 cm$^{-1}$. Narrowband light is generated by combining a 40 cm$^{-1}$ broad 1030 nm pump beam with a 250 cm$^{-1}$ broad tunable idler beam in a 1.5 mm thick GaSe crystal. The full width at half maximum spectral width of the narrowband channel is reduced to around 30 cm$^{-1}$ (red trace). We then send the mid-infrared light to a home-built monochromator by placing a slit in the Fourier plane of a pulse shaper (composed of a grating, followed by a cylindrical lens, a slit, another cylindrical lens, and another grating) to select radiation within a narrow bandwidth <4 cm$^{-1}$ (*44*). The result is a channel of ultra-narrowband radiation with spectral width <4cm$^{-1}$ (blue trace). We remark that the spectrometer used to measure the data in Fig. S14 has a resolution of only around 4 cm$^{-1}$, and thus the ultra-narrowband radiation could have a linewidth narrower than 4 cm$^{-1}$.

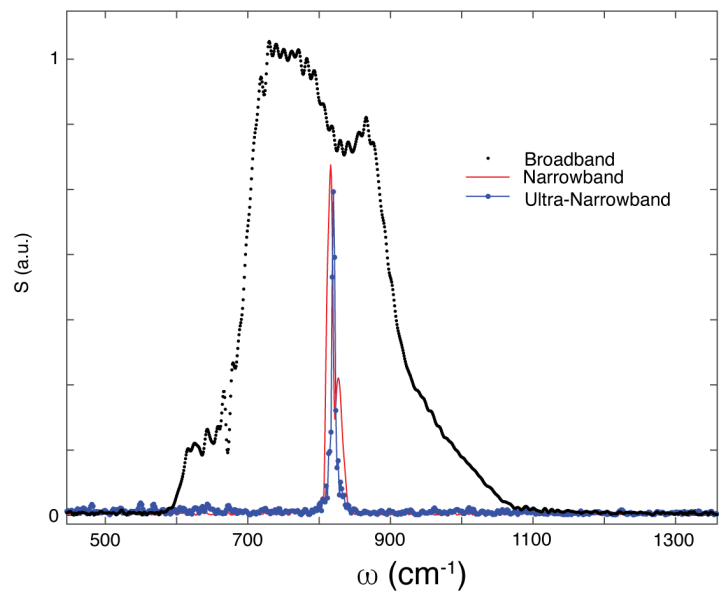

Figure S14| **Spectra of the infrared light sources used in this work.** (see text)

*S2.5. Fundamental limits of the polariton width*

To address the important challenge of focusing hyperbolic polaritons (*18, 32, 33*), we discuss factors that limit the polariton's width in this section. Representative data obtained on two hetero-bicrystals at the frequency $\omega_0$ of two different thicknesses are shown in Fig. S15A and S15C. We observe maxima in the near-field amplitude at positions that are in lateral alignment with the edges of gold strips, buried beneath the hetero-bicrystals. The full-width-at-half-maximum of these peaks relative to the baseline intensity at the center of the launcher, *D*, are *D* = 140 nm for the data in Fig. S15A and *D* = 90 nm for the data in Fig. S15C. The general trends observed in the data are captured with numerical solutions to Eq. (S30) displayed in Fig. S15B and S15D. We emphasize that the width increases with increasing bicrystal thickness in both theory and experiment. However, critically, to account for the experimentally observed widths, it was necessary to include a large air gap between the gold launcher and the sample, $h = 20\ nm$, in our simulations (see inset in Fig. S15G).

To test the theoretical limits of the attainable spot size we carried out numerical simulations shown in Fig. S15E-G. We display the full-width-at-half-maximum of the observed peaks near the edges of the launchers, as schematically shown in Fig. S15B and S15D. The results in Fig. S15E show that the width, *D*, decreases as the bicrystal thickness is reduced, while other parameters are held fixed. Further, the results shown in Fig. S15F show that the width can theoretically decrease for a bicrystal of a given thickness if the losses of the crystals are reduced. These results in Fig. S15F were obtained by multiplying the imaginary components of the permittivity of both crystal by an artificial "damping factor" while all other parameters in the simulation were fixed. Finally, in Fig. S15G we show the spot size increases as $h$ increases, e.g. as the quality of the launcher worsens, with the other parameters held constant.

These results are all consistent with the relationship $w \propto Im(\delta_{tot}) + 2h$ given in Ref (*40*). The displacement of the ray, $\delta_{tot} = \delta_A + \delta_B$ (see Eq. (14)) is a complex quantity with an imaginary component that depends on the thicknesses of crystals "A" and "B" as well as the linewidth of their phonon resonances. Thus, the width can be considerably sharpened by reducing the bicrystal thickness, even with fixed losses. If the losses can be reduced, either by decreasing the sample's temperature or improving linewidths of the phonons (or generally the dipole active resonances responsible for hyperbolicity), the polariton's spot-size could theoretically be further sharpened. However, the quality of the launcher can also impose limitations on the attainable spot sizes. Our simulations support the notion that $h \sim 20\ nm$ is necessary to account for the width in the present experiments, which limits the attainable spot size. Since the experiments were performed with relatively thick launchers with imperfect edges, it may be possible to reduce the width, *D*, by optimizing the quality of the launchers in future experiments.

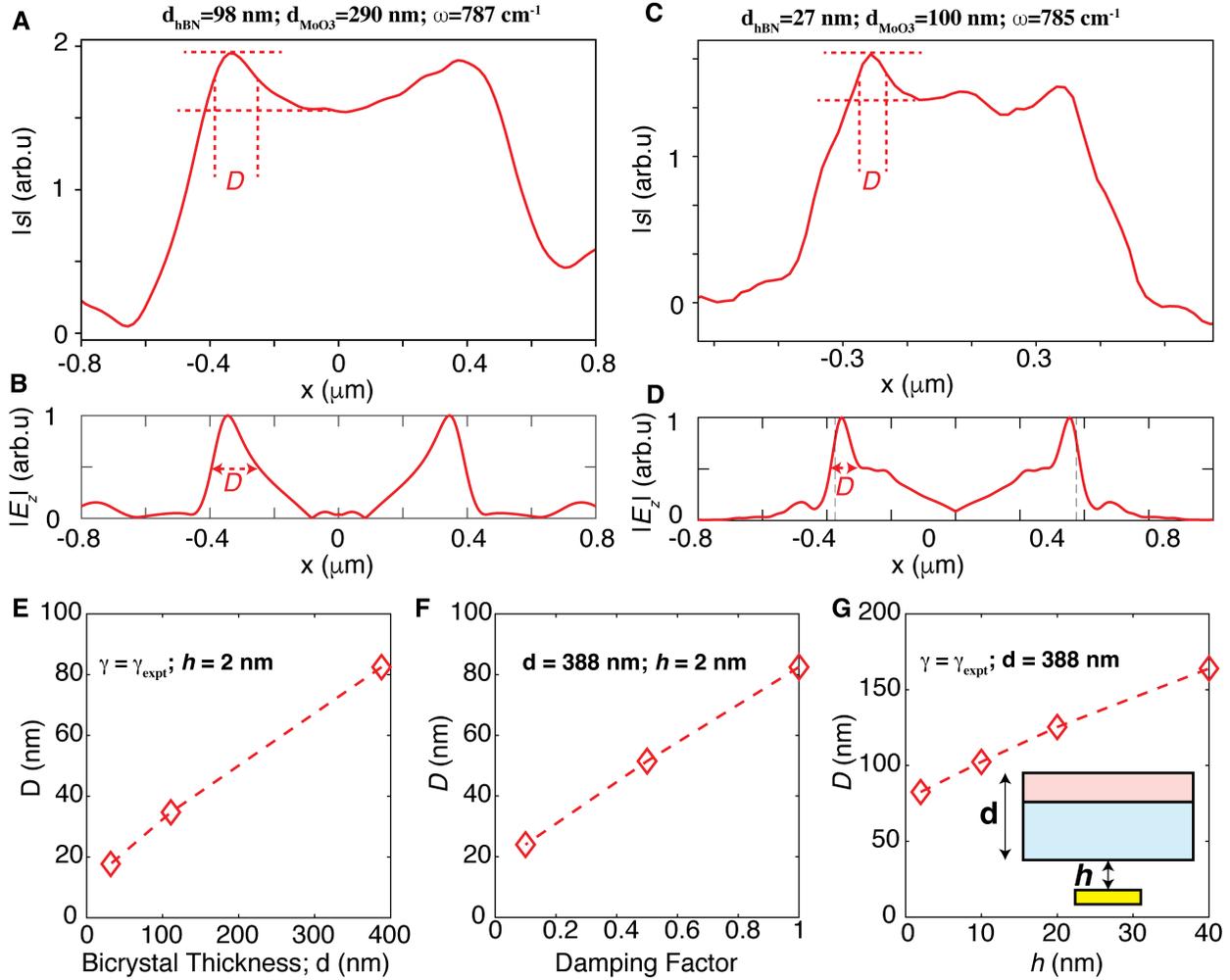

Figure S15| **Fundamental limits of the spot-size. A** and **C**, Data obtained on (**A**) a heterobicrystal with $d_{hBN}$=98 nm and $d_{MoO3}$ = 290 nm at the frequency ω=787 cm$^{-1}$ and (**C**) a heterobicrystal with $d_{hBN}$ = 27 nm and $d_{MoO3}$ = 100 nm at the frequency ω = 785 cm$^{-1}$. **B**, calculations using $h = 20\ nm$ and the thickness and frequency parameters corresponding to panel (**A**). **D**, same as panel (**B**) with thickness and frequency parameters corresponding to panel (**C**). **E**-**G** Calculations of the full-width-at-half maximum of the peak '*D*', indicated in panels (**B**) and (**D**), for a series of parameters. Panel **E**, the damping is equal to the experimental value and a small air gap, $h = 2\ nm$, is considered while the total thickness of the hetero-bicrystal is varied (with the ratio $d_{MoO3} \cong 3\ d_{hBN}$ fixed). Panel **F**, calculations with fixed values of d=388 nm and $h = 2nm$, while artificially reducing losses with a 'damping factor' (see text). Panel **G**, the damping, equal to the experimental value, and d = 388 nm are fixed while $h$ is varied.